\documentclass[%
reprint,
nofootinbib,
amsmath,
amssymb,
prd,
longbibliography,
superscriptaddress]{revtex4-2}

\usepackage{mathrsfs}
\usepackage{graphicx}
\usepackage{dcolumn}
\usepackage{bm}
\usepackage{bbm}
\usepackage{xcolor}
\usepackage{microtype}
\usepackage{IEEEtrantools}
\usepackage{amsthm}
\usepackage{enumitem}
\usepackage[normalem]{ulem}
\usepackage{multirow}
\usepackage{hyperref}
\hypersetup{breaklinks=true}
\usepackage[T1]{fontenc}

\usepackage[capitalise]{cleveref}

\begin{document}

\title{Extreme-mass-ratio inspirals in relativistic accretion discs}
\author{Francisco Duque}
\affiliation{Max Planck Institute for Gravitational Physics (Albert Einstein Institute) 
Am Mühlenberg 1, D-14476 Potsdam, Germany}
\author{Laura Sberna}
\affiliation{School of Mathematical Sciences, University of Nottingham, University Park, Nottingham NG7 2RD, United Kingdom}
\author{Andrew Spiers}
\affiliation{School of Mathematical Sciences, University of Nottingham, University Park, Nottingham NG7 2RD, United Kingdom}
\author{Rodrigo Vicente}
\affiliation{Gravitation Astroparticle Physics Amsterdam (GRAPPA), University of Amsterdam, Science Park 904, 1098 XH, Amsterdam, The Netherlands}
\date{\today}

\begin{abstract}
We compute relativistic Lindblad torques for circular, equatorial extreme-mass-ratio inspirals (EMRIs) embedded in relativistic thin accretion discs, including spinning black hole configurations. We find that relativistic effects can amplify the magnitude of these torques by orders of magnitude in the strong-field regime, and that the torque can even reverse direction as the EMRI approaches the innermost stable circular orbit (ISCO). However, we show that the location of this reversal is highly spin-dependent, shifting progressively closer to the ISCO, where gravitational-wave emission completely dominates the inspiral, as the spin of the central black hole increases. Spin also modifies the radial dependence of the Lindblad torques. We investigate whether Lindblad torques can be approximated by parametrised power laws of the form $T_\text{LR} = A \left( r_s/10M \right)^{n_r}$
(or combinations thereof), and find significant spin- and disc-dependent variations in the slope parameter $n_r$. For instance, for spin $a/M = 0.9$, we find $n_r = 3.6$ in the strong-field regime, compared to the Newtonian value of $n_r = 4.5$. Given current forecasts of parameter recovery for ``golden'', loud EMRIs in accretion discs ($\Delta n_r \sim 0.5$), we predict LISA could distinguish between different disc configurations through their relativistic Lindblad torque signatures, providing the first direct probe of the midplane structure of the inner region of accretion discs, which is inaccessible to electromagnetic observations.
\end{abstract}

\maketitle

\section{Introduction}

Dense astrophysical environments at the centres of galaxies---such as nuclear clusters~\cite{Neumayer:2020gno, Miller:2008yw}, the discs of active galactic nuclei (AGN)~\cite{Abramowicz:2011xu}, and dark-matter overdensities~\cite{Gondolo:1999ef, Brito:2015oca, Hui:2021tkt}---have assumed growing interest in gravitational-wave (GW) astronomy. It is now well established that these structures can significantly influence the formation, evolution, and GW signature of compact binaries across the entire GW spectrum~\cite{Kocsis:2011dr, Barausse:2014pra, Speri:2022upm,  Cole:2022yzw, NANOGrav:2024nmo, Garg:2022nko, CanevaSantoro:2023aol, Roy:2024rhe, Zwick:2025wkt, Tomaselli:2024ojz, Chen:2025qyj, Zwick:2021dlg, Derdzinski:2020wlw, Santos:2025ass, Chen:2025qyj, Hu:2023oiu}. This increased activity is partially driven by the recognition that relativistic effects are crucial to accurately model the interaction between binaries and their astrophysical surroundings~\cite{Barausse:2007ph, Cardoso:2021wlq, Vicente:2025gsg, Khalvati:2024tzz, Brito:2023pyl, Cardoso:2022whc, Datta:2021hvm, Duque:2023cac, dyson2025environmental, Polcar:2025yto, Xin:2025ymm, Guo:2025pea, Aurrekoetxea:2024cqd}.

Extreme-mass-ratio inspirals (EMRIs) are particularly promising candidates to detect environmental effects in GWs. EMRIs consist of a \textit{secondary} stellar-mass compact object---typically a black hole (BH) with mass $\mu \sim 10$--$100\, M_\odot$---slowly inspiralling into a \textit{primary} massive black hole (MBH) with $M \gtrsim 10^5\, M_\odot$ at the centre of a galaxy~\cite{Babak:2017tow, Mancieri:2025cmx}. EMRIs are unique probes of strong-field gravity, as they execute tens of thousands of relativistic orbits while in the frequency band accessible to the Laser Interferometer Space Antenna (LISA), a space-based GW observatory planned for launch in 2035~\cite{Colpi:2024xhw}, as well as to the proposed missions Taiji and TianQin \cite{Gong:2021gvw}. The highly asymmetric mass ratio in EMRIs renders standard techniques developed for comparable-mass binaries inadequate. Instead, the relativistic two-body problem is addressed by the self-force program, which employs BH perturbation theory to expand the Einstein equations in the small mass ratio $q = \mu / M$~\cite{Pound:2021qin, wardellpoundreview2021}.

To extend the self-force framework to non-vacuum backgrounds, one can introduce an additional expansion parameter encoding the energy density of the surrounding medium~\cite{Cardoso:2022whc, Datta:2021hvm, Duque:2023cac, dyson2025environmental, Polcar:2025yto}. The Einstein equations are then solved self-consistently alongside the equations governing the matter fields, enabling the systematic inclusion of processes such as accretion, dynamical friction, and back-reaction on the environment. Applications of this method to EMRIs in dark-matter overdensities have revealed that relativistic corrections substantially strengthen binary–environment interactions compared to Newtonian predictions~\cite{Duque:2023cac, dyson2025environmental, Vicente:2025gsg}. 

The relativistic modeling of EMRIs in accretion discs remains far less developed. This is primarily due to the complexity of modelling strong gravity together with the challenging microphysics of discs, including radiation transport, viscosity, and magnetic fields~\cite{Abramowicz:2011xu}. Nonetheless, EMRIs in AGN discs represent a compelling science case for LISA, offering a potential window into accretion processes near MBHs~\cite{Speri:2022upm, Duque:2024mfw, Zwick:2021dlg, Khalvati:2024tzz, Chen:2025qyj, Derdzinski:2020wlw, Sun:2025lbr}. Intriguingly, indirect evidence for such systems may already exist in the form of quasi-periodic eruptions (QPEs)---sharp, recurrent X-ray bursts observed in galactic centres on timescales of hours to days. QPEs have been interpreted as signatures of a compact object periodically crossing an accretion disc, exciting shocks or density perturbations that modulate the X-ray flux~\cite{Franchini:2023bou, Linial:2023nqs, Chakraborty:2025xch}. 

A key property of accretion discs is their differential rotation. At specific radii, particles within the disc can orbit with frequencies that are integer multiples of the time-varying gravitational potential induced by the EMRI. These resonant conditions enable an efficient exchange of angular momentum, giving rise to Lindblad resonances, whose associated torques play a central role in shaping the coupled evolution of the system.
Lindblad torques have been extensively investigated in Newtonian gravity, particularly in the context of disc–planet interactions~\cite{GoldreichTremaine1980, Artymowicz1993, Tanaka_2002, Tanaka_2004, Tanaka3, rafikov_eccentricity}. In such systems, the net torque generated by Lindblad resonances drives the evolution of the planet’s orbit, causing migration either inward or outward depending on the relative strengths of the inner and outer Lindblad torques. These Newtonian results have been directly applied to the study of EMRIs embedded in accretion discs. It was shown that they are the dominant environmental effect for these systems~\cite{Kocsis:2011dr}, and they could be measurable by LISA~\cite{Yunes:2011ws, Speri:2022upm, Khalvati:2024tzz}. Neglecting these torques in waveform modelling can lead to substantial biases in parameter estimation, introducing systematic errors, causing signals to be missed or misinterpreted as deviations from General Relativity (GR)~\cite{Speri:2022upm, Kejriwal:2023djc}.

Earlier studies already suggested that accurate EMRI waveform templates would require relativistic models. In 2011, Hirata~\cite{Hirata:2010vn, Hirata:2010vp} (henceforth referred to as Hirata I~\cite{Hirata:2010vn} and Hirata II~\cite{Hirata:2010vp}) extended the computation of Lindblad torques into the fully relativistic regime, calculating Lindblad resonances as a perturbative correction to the Kerr spacetime surrounded by an arbitrarily thin disc.\footnote{The series' first paper~\cite{Hirata:2010vn} does not specify that the primary BH is Kerr, or the disc being thin; it simply restricts to axisymmetric, time-stationary spacetime with a plane of symmetry.} The disc is assumed to be a two-dimensional, \textit{pressureless} collection of particles in circular geodesics around the primary MBH. Hirata's analyses demonstrated that general relativistic effects systematically enhance the magnitude of these torques compared to Newtonian predictions. However, this pioneering work focused primarily on the formalism and only calculated the torque mode by mode, normalised by the local disc density, and did not explore the implications for GW observations of EMRIs. Another limitation of Hirata's model is that it does not include disc pressure, which is responsible for maintaining the disc's vertical structure and propagating density waves in the disc. As we will see, without pressure corrections, the total Linbdlad torque exerted on the EMRI diverges. 

These gaps motivate our study. Building on Hirata’s framework, we compute the total relativistic Lindblad torque on an EMRI in a relativistic accretion disc and regularise it with a phenomenological pressure prescription, inspired by Newtonian migration theory. The paper is structured as follows. Sec. \ref{sec:Summary} summarises the three main conclusions of our work. Sec.~\ref{sec:formalism} reviews Hirata’s formalism and highlights the divergence of mode sums without pressure. Sec.~\ref{sec:TorquePressure} introduces our pressure regularization and applies it to several disc models, yielding the results of Fig.~\ref{fig:TorqueGWComp}. We conclude with implications for GW astronomy. Unless otherwise stated, we adopt $G = c = 1$, and use primes to denote radial derivatives.

\section{Executive Summary}\label{sec:Summary}

\begin{figure}[t]
\includegraphics[width=0.995\linewidth]{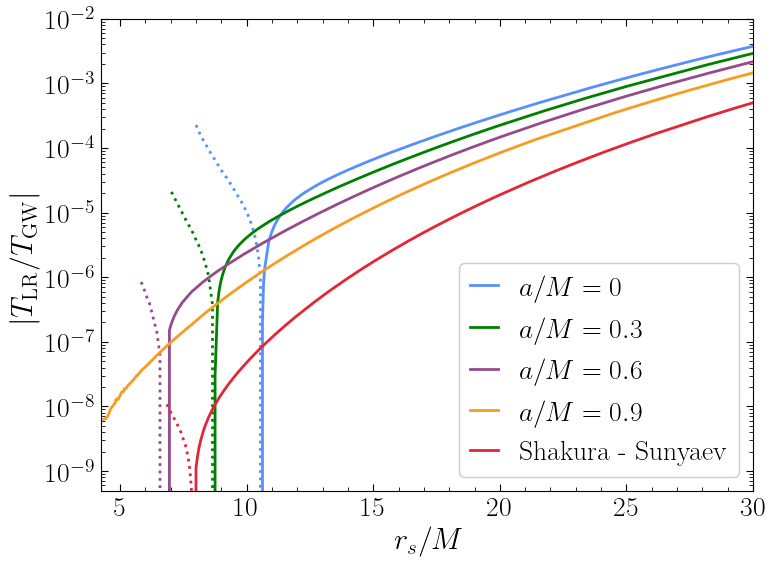}   
\caption{Ratio of the relativistic Lindblad torques to the gravitational-wave torques for an equatorial, circular EMRI embedded in different disc configurations, shown as a function of orbital separation $r_s$ and for representative disc models with the same viscosity and accretion rate (see Fig.~\ref{fig:discProfile}). For spinning primaries, we use the relativistic Novikov-Thorne disc model. Dotted segments indicate regions where the Lindblad torque becomes negative.}
\label{fig:TorqueGWComp}
\end{figure}

Our main results are summarised in Fig.~\ref{fig:TorqueGWComp}, which compares the relativistic and Newtonian torque strengths across different disc models (the relativistic Novikov–Thorne profiles at various spins, and the Newtonian Shakura–Sunyaev profile, with the same accretion rate and viscosity). 

We highlight three conclusions:
\begin{enumerate}
    \item Relativistic effects substantially amplify accretion torques. Both relativistic corrections to the binary–disc interaction and to the disc structure itself can increase torque amplitudes by orders of magnitude;
    \item The torque scaling with orbital radius is altered. Lindblad torques can still be described by power laws of the form $T_\text{LR} = A (r_s/r_0)^{n_r}$, but the slope $n_r$ is spin-dependent and consistently smaller than the Newtonian prediction, particularly in the strong-field regime, implying Lindblad torques are more relevant for small orbital separations than previously expected (see Fig.~\ref{fig:TorqueStrengths}); 
    \item Near the ISCO, torques can reverse sign (shown by the dotted lines in Fig.~\ref{fig:TorqueGWComp}). This reversal is associated with the accumulation of inner Lindblad resonances closer to the secondary than the outer resonances as the orbit approaches the ISCO. The reversal radius depends on the MBH spin, occurring closer to the ISCO for higher spins.
\end{enumerate}

As we were finalizing this work, we became aware of independent efforts~\cite{HegadeKR:2025dur, HegadeKR:2025rpr} pursuing the same problem through a (semi-)analytical Hamiltonian approach, improving on Hirata’s original formulation. The methodologies differ: our work focuses on numerically evaluating small multipoles and extrapolating to large ones, whereas Refs.~\cite{HegadeKR:2025dur, HegadeKR:2025rpr} derive analytical expressions at large multipoles and extrapolates downward. Our findings are consistent, reinforcing the robustness of these results.

%

\section{Relativistic Lindblad torques in pressureless 2D discs}\label{sec:formalism}

We start by summarising Hirata's seminal results on relativistic Lindblad resonances for pressureless 2D discs in axisymmetric and stationary spacetimes. Throughout, we assume that the disc is aligned with the spin of the central BH and neglect its backreaction on the Kerr geometry. Both assumptions are well justified in the inner disc region, where alignment is expected from the Bardeen–Petterson or magnetic spin–alignment mechanisms~\cite{Bardeen:1975zz, Natarajan:1998xt, Chatterjee:2023ber}, and the disc’s self-gravity is negligible~\cite{Lobban:2022aon}. Additionally, we focus on \emph{prograde} EMRI circular orbits, since interactions with the disc are expected to align their angular momentum orientation and damp eccentricity before the inspiral reaches the inner regions relevant to the LISA band (although residual eccentricity could survive in-band~\cite{Spieksma:2025wex, 2024MNRAS.528.4958W}). The treatment can be readily extended to EMRIs on eccentric orbits. 
We refer the reader to Refs.~\cite{Hirata:2010vn,Hirata:2010vp} for complete details.

\subsection{Location of the resonances} \label{sec:ResLocation}
%
\begin{figure*}[t]
\includegraphics[width=0.495\linewidth]{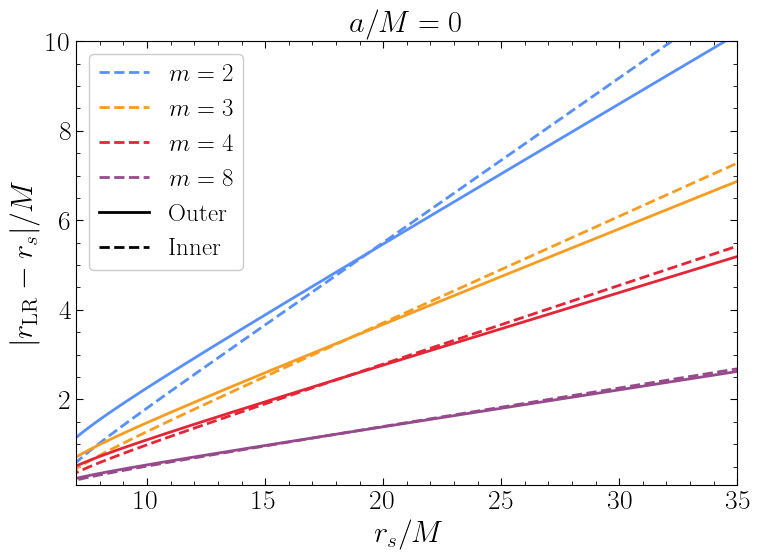}    
\includegraphics[width=0.495\linewidth]{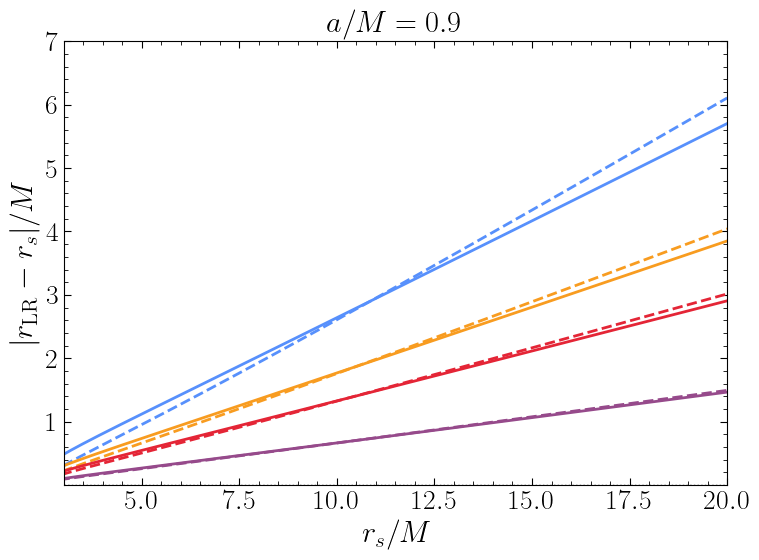}
\caption{Radial location of Lindblad resonances, $r_\text{LR}$, as a function of the secondary orbital radius, $r_s$, for different azimuthal mode numbers $m$ and primary BH spin $a$. In the large-$m$ limit, the inner and outer Lindblad resonances become equidistant and close to the secondary. For large orbital radii, outer Lindblad resonances are consistently located closer to the secondary than their inner counterparts (for the same $m$), but this hierarchy reverses at smaller $r_s$. The critical radius at which this switch occurs is spin-dependent, decreasing with increasing BH spin parameter $a$.}
\label{fig:LRlocations}
\end{figure*}

Let us start by discussing the radial location of the resonance in the disc for a given radius of the circular orbit of the secondary. Assuming that disc particles follow approximately circular geodesics in the background spacetime, a particle that is radially displaced will undergo oscillations about its original orbit at a frequency referred to as the (radial) epicyclic frequency. In the limit of small eccentricities, this epicyclic frequency coincides with the radial orbital frequency.
Lindblad resonances arise at locations where the epicyclic frequency of the disc is commensurate with the perturbation frequency from the secondary object. More precisely, these resonances can be identified by the vanishing of the \textit{resonant detuning function}
\begin{align}\label{eq:detuningfunc}
    D[r]:= m(\Omega[r]-\Omega_s)\mp \kappa[r] \, , \quad  m \in \mathbb{N}
\end{align}
where the subscript $s$ refers to the secondary, $\Omega$ is the orbital frequency for circular equatorial geodesics in Kerr 
\begin{align}
    \Omega[r] = \frac{M^{1/2}}{r^{3/2}+ M^{1/2} a} \, ,
\end{align}
with $a$ the spin parameter, and $\kappa$ is the radial epicyclic frequency in Kerr,
\begin{align}
    \kappa[r] = \Omega[r]\sqrt{1 - \frac{6M}{r} + \frac{8 a M^{1/2}}{r^{3/2}} -  \frac{3a^2}{r^2}}\, .
\end{align}
The $-$ sign version of the detuning function identifies inner (with respect to the secondary) Lindblad resonances, whereas the $+$ version identifies outer Lindblad resonances. For any given $\Omega_s$, there are infinitely many inner and outer Lindblad resonance positions ($r_\text{ILR}$ and $r_\text{OLR}$ respectively) that satisfy $D[r_\text{LR}]=0$, one for each value of $m$. Outer Lindblad resonances are responsible for inward migration, while inner ones lead to outward migration. 

In Fig.~\ref{fig:LRlocations}, we plot the locations of Lindblad resonances as a function of the secondary’s orbital radius $r_s$, for various values of $m$ and primary spin parameter $a$. In the Newtonian regime (large $r_s$), the outer Lindblad resonances lie closer to the secondary than the inner ones. For a disc with a constant surface density profile, this naturally causes the outer resonances to dominate at large radii, resulting in a net inward Lindblad torque on the EMRI---thus accelerating its inspiral compared to one driven by vacuum GW emission.

As the secondary moves inward and $r_s$ decreases, this pattern reverses: the inner Lindblad resonances move closer to the secondary than the outer ones, and the net disc torque may become outward. If GW emission dominates the evolution, this slows down the inspiral with respect to vacuum. If migration torques compete with or dominate over GW emission, the inspiral can be stalled, forming a migration trap for the EMRI. The critical radius where the inner and outer resonance locations switch proximity is spin-dependent---it occurs at smaller $r_s$ for higher $a$. Specifically, for a non-spinning primary, $a = 0$, this transition happens at $r_s \sim 18M$, while for a rapidly spinning one, $a = 0.9M$, it is shifted inward to approximately $r_s \sim 10M$. This feature is also present in the analysis of Refs.~\cite{HegadeKR:2025dur, HegadeKR:2025rpr}.

This switch radius also depends on the resonance index $m$: it shifts to smaller values for larger $m$, but saturates in the $m \rightarrow \infty$ limit, with the inner and outer Lindblad resonances becoming equidistant from the secondary. Note also that the inner resonances never fall below the ISCO. Additionally, higher-$m$ resonances always lie closer to the secondary, which causes them to contribute more strongly to the total Lindblad torque (when summing over all $m$). Naively, this sum would diverge, but as we will discuss in later sections, it is regularized by including disc pressure effects, which were absent in Hirata's original treatment.


\subsection{Mode by mode torque amplitudes}



In \cite{Hirata:2010vn,Hirata:2010vp}, Hirata developed a relativistic framework for computing Lindblad torques based on the Hamiltonian formalism. Remarkably, this approach demonstrates that the change of energy and angular momentum of the secondary can be expressed in terms of the power emitted (to future null infinity and the future horizon) by the interference between GWs emitted by the secondary and those emitted by a disc particle in a slightly eccentric orbit. The interference is constructive only for particles located at Lindblad resonance radii, where particles emit GWs coherently with the perturber undergoing circular motion. Consequently, the Lindblad torques can be directly calculated from the same gauge-invariant quantities that govern GW emission in EMRIs---specifically, from solutions of the Teukolsky equation~\cite{teukolsky1974perturbations}.

\subsubsection{Definitions: the Kerr metric, circular geodesics, and Teukolsky formalism}

To make this section self-contained, we introduce definitions of the Kerr metric~\cite{kerr1963gravitational} and the Teukolsky formalism~\cite{teuk1972,teuk1973}, which will be used in Hirata's relativistic Lindblad resonance formalism. For comprehensive reviews, see Refs.~\cite{chandrabook,wardellpoundreview2021}. The explicit form of the line element of the Kerr metric in Boyer-Lindquist coordinates $(t, \, r, \, \vartheta, \, \varphi)$ is
\begin{align}
&ds^2 =  -\left( 1 - \frac{2 M r}{|\rho|^2} \right) dt^2 - \frac{4 M a r}{|\rho|^2} \sin^2\vartheta \, dt \, d\varphi
\nonumber \\ 
& + \frac{|\rho|^2}{\Delta} dr^2 + |\rho|^2 d\vartheta^2
+ \frac{(r^2 + a^2)^2 - \Delta a^2 \sin^2 \vartheta}{|\rho|^2} \sin^2\vartheta\, d\varphi^2  \,  ,
\end{align}
where
\begin{align}
\Delta = r^2 - 2 M r + a^2\, , \quad  |\rho|^2 = r^2 + a^2 \cos^2\vartheta.
\end{align}
From this, one can obtain the contravariant metric coefficients
\begin{align}
g^{tt}      &= - \frac{(r^2+a^2)^2 - \Delta a^2 \sin^2\vartheta}{\Delta|\rho|^2}\, , \\[1em]
g^{t\varphi}   &= - \frac{2a M r}{\Delta|\rho|^2}\, , \quad
g^{\varphi\varphi}= \frac{\Delta - a^2 \sin^2\vartheta}{\Delta|\rho|^2\sin^2\vartheta}\,, \\[1em]
g^{rr}      &= \frac{\Delta}{|\rho|^2}\,, \qquad
g^{\vartheta\vartheta} = \frac{1}{|\rho|^2}. 
\end{align}

Equatorial circular geodesics in Kerr spacetime are described by the four velocity $w^\alpha$, with
\begin{align}
    w^t &= \frac{dt}{d\tau_s} = -g^{tt} E + g^{t\varphi} L_z \, ,\\
     w^r &=0,\\
     w^\theta &=0,\\
     w^\phi &=\frac{d\phi}{d\tau_s} =-g^{t\phi}E + g^{\phi\phi}L_z   \, , 
\end{align} 
where $\tau_s$ is the secondary's proper time, and $E$ and $L_z$ are, respectively, its energy and angular momentum, given by
\begin{align}
    E &= 
\frac{ r^{3/2} - 2 M r^{1/2} + a M^{1/2} }
     { r^{3/4} \, \sqrt{ r^{3/2} - 3 M r^{1/2} + 2 a M^{1/2} } } \, , \\
 L_z &= \frac{M^{1/2} r^{2} - a M^{1/2} (2r - M)}{r^{3/2} - 2 M r^{1/2} + a M^{1/2}} \, .    
\end{align}

The emission of GWs by an EMRI in Kerr (to first-order in the mass ratio) is governed by the Teukolsky equation for the Newman-Penrose Weyl scalar $\psi_{4}$~\cite{teukolsky1974perturbations}
\begin{align}
   _{-2} \mathcal{O}[\psi_4] = T \, , 
\end{align}
where $_{-2} \mathcal{O}$ is a second-order differential operator and $T$ represents a point-particle source---the secondary or a particle in the disc---that scales linearly with the mass-ratio $q=\mu/M$. $\psi_4$ is directly related to the gravitational wave polarizations at infinity,
\begin{align}
   \psi_4 \to  \frac{1}{2} \frac{d^2}{dt^2}\left( h_+ - i h_\times \right) \qquad \text{as} \, \, r\rightarrow \infty \, .
\end{align}
The Teukolsky equation is separable in the frequency domain with the decomposition
\begin{align}
\psi_4 &= \frac{1}{\rho^4} 
\int_{-\infty}^{\infty} d\omega 
\sum_{\ell=2}^{\infty} \sum_{m=-\ell}^{\ell} 
R_{\ell  m}(r; \omega)\, _{-2} S_{\ell m}(\vartheta; \chi ) \,
e^{i \left( m\varphi - \omega t  \right)} \, ,
\end{align}
where
\begin{align}
    \chi = a \omega \, ,   \quad \rho = r - i a \cos \vartheta \, .
\end{align}
Here,  $_{-2} S_{\ell m}$ are the spin-weighted spheroidal harmonics with spin-weight $s=-2$, which satisfy the eigenvalue equation
\begin{equation}
    _{-2}\mathcal{O}_{\vartheta\varphi} \left[ _{-2} S_{\ell m} \right] = \,   _{-2}\mathcal{E}_{\ell m} \,  _{-2} S_{\ell m} \, , 
\end{equation}
where  $_{-2}\mathcal{O}_{\vartheta\varphi}$ is a second-order differential operator containing only derivatives in $\vartheta$ and $\varphi$, while $_{-2}\mathcal{E}_{\ell m}$ are the corresponding eigenvalues, which in Kerr are frequency dependent.  

The radial function $R_{\ell  m}$ satisfies the inhomogeneous radial Teukolsky equation
\begin{align}
\Delta^2 \frac{d}{dr}\left(\Delta^{-1}\frac{dR}{dr}\right) - VR = -\mathcal{T} \, , \label{eq:Radial}
\end{align}
where $\mathcal{T}$ represents the source term, $V = V\left(r ; \, \omega, \, \lambda_{\ell m } \right)$ is the effective potential, and  
\begin{align}
    \lambda_{\ell m} = \, _{-2}\mathcal{E}_{\ell m} - 2 m \varpi + \varpi^2 -2 \, , 
\end{align}
with the frame-dragging shifted frequency
\begin{align} 
    \varpi = \omega - m\,\Omega_\text{H} \, ,
\end{align}
where $\Omega_\text{H}$ is the angular velocity of the horizon 
\begin{align}
\Omega_\text{H} = \frac{a}{2 r_+} \, ,
\end{align}
and $r_+$ is the event horizon radius
\begin{align}
r_{+} = M + \sqrt{M^{2} - a^{2}} \, .
\end{align}

The physical solutions of Eq.~\eqref{eq:Radial} correspond to outgoing waves at infinity and ingoing ones at the horizon
\begin{align}
R^\infty_{\ell m}(r) &\to 
    Z^{\infty}_{\ell m \omega}\, r^{3} e^{i \omega r_*} \,,
    & r \rightarrow \infty \,, \label{eq:OutgoingWave}  \\[0.5em]
R^\text{H}_{\ell m}(r) &\to 
    Z^{\mathrm{H}}_{\ell m\omega}\, \Delta\, 
    e^{-i \varpi r_*} \,,
    & r \rightarrow r_+ \,, \label{eq:IngoingWave} 
\end{align}
where $r_*$ is the tortoise radial coordinate
\begin{align}
    \frac{dr_*}{dr} = \frac{r^2 + a^2}{\Delta } \, . 
\end{align}

For generic EMRIs (inclined and eccentric orbits), the amplitudes $Z^{\infty, \, \text{H}}_{\ell m \omega}$ admit a multi-modal decomposition~\cite{Hughes:2021exa}
\begin{align}
  Z^{\infty, \, \text{H}}_{\ell m \omega} = \sum_{k = -\infty }^{\infty}\sum_{n = -\infty }^{\infty}   Z^{\infty, \, \text{H}}_{\ell m n k} \delta(\omega - \omega_{m n k} )\, , 
\end{align}
where
\begin{align}
 \omega_{m n k} = m \Omega_\varphi + n \Omega_r + k \Omega_\vartheta \, , 
\end{align}
and where $\Omega_\varphi, \, \Omega_r,  \, \Omega_\vartheta $ are the fundamental frequencies for bound motion in the azimuthal, radial, and polar direction, respectively. For equatorial orbits, there is no motion in $\vartheta$ and thus we set $k = 0$. Similarly, for circular orbits $n=0$. The power emitted in GWs to infinity relates to the Teukolsky amplitudes by~\cite{teuk1972, teuk1973}
\begin{align}\label{eq:P_infinity}
P_\infty =\sum_{\ell, m,  n}  \frac{| Z^\infty_{\ell m n}|^2}{2 \omega_{mn}^2} \, ; 
\end{align}
and the power absorbed at the horizon is 
\begin{align}
P_\text{H} =\sum_{\ell, m,  n}  \alpha \frac{|Z^\text{H}_{\ell m n}| ^2}{2 \omega_{mn}^2} \, , 
\end{align}
where 
\begin{align}
\alpha &=
\frac{
8192\, r_{+}\; \varpi \;\left( \varpi^{2} + \Gamma^{2} \right) \left( \varpi^{2} + 4\,\Gamma^{2} \right) \omega^{3}
}{
|C|^{2}  
} \, ,\\ 
\Gamma& = \frac{2\,\sqrt{\,1 - a^{2}\,}}{a^{2} + r_{+}^{2}} \, , \\
|C|^{2} &= \left[ \left( \lambda_{\ell m} + 2 \right)^{2} + 4 m \varpi - 4 \varpi^{2} \right] \big[ \lambda_{\ell m}^{2} + 36 m \varpi - 36 \varpi^{2} \big] \nonumber \\
&+ 48( 2\lambda_{\ell m} + 3 )\,\varpi\,( 2\varpi - m ) + 144\,\omega^{2}\,(1 - a^{2}) \, .
\end{align}

Hirata exploited a known relation~\cite{} between ingoing and outgoing Teukolsky amplitudes, reducing the number of amplitudes that must be computed. The outgoing solution, Eq.~\eqref{eq:OutgoingWave}, can be expressed in terms of its near-horizon amplitude via connection coefficients $c_{13}$ and $c_{14}$:
\begin{align}
    R^\infty_{\ell m}(r)&\to  Z^{\mathrm{H}}_{\ell m\omega} \left(c_{13} r^3 e^{i \omega r_*} + c_{14} \frac{ e^{-i \omega r_*} }{r}  \right) \, ,  & r \rightarrow \infty \, ,
\end{align}
which are complex constants that can be determined numerically. The power of the outgoing wave, \cref{eq:P_infinity}, is then
\begin{align}
P_\infty &= \sum_{\ell, m} \frac{\left|c_{13} Z^\text{H}_{\ell m \omega} \right|^2}{2 \omega^2} \, . 
\end{align}

\subsubsection{Relativistic Lindblad torques formalism}

The Hirata's Lindblad torque strength is given by~\cite{Hirata:2010vn} 
\begin{equation}\label{eq:TLRm}
T^{(m)}_\text{LR} = \mp \frac{2\pi^{2} m }{\left|D'(r_\text{LR})\right|}\,
r_\text{LR}\, w^{t}\, \mathcal{Z}\, \Sigma \,\big|S^{(m)}\big|^{2}\,,
\end{equation}
where all quantities are evaluated at the resonance $r_\text{LR}$, and $\mathcal{Z}$ is the specific epicyclic impedance, 
\begin{align}
    \mathcal{Z} =  \sqrt{\frac{C_{020}}{C_{002}} } \Bigg |_{r=r_\text{LR}}  \, ,
\end{align}
with 
\begin{align}
   C_{002} &= \frac{g^{rr}}{w^t} \, , \\
   C_{020} &= g^{\varphi \varphi}\left(\frac{g^{\varphi \varphi} - g^{tt}}{(w^t)^3}\right)\left( \frac{dL_z}{dr} \right)^2 - \left( \frac{d\Omega}{dr} \right)\left( \frac{dL_z}{dr} \right) \, .
\end{align}
Here, $\Sigma$ is the surface density of the 2D disc (at the resonance), and $\mathcal{S}^{(m)}$ are the resonant amplitudes.\footnote{Our definition of $T^{(m)}_\text{LR}$ is derived from Hirata I Eq.~(92) integrated against $r$; this should not be confused with $T$ in Hirata I Eq.~(92).}

The resonant amplitudes can be related to a power amplitude~\cite{Hirata:2010vp}, $\mathcal{P}_\text{LR}^{(m)}$,
\begin{align}\label{eq:torqueSm}
\mathcal{S}^{(m)} = \lim_{e\to 0} \frac{2i}{m \Omega_s \mu \mathcal {Z}_\text{LR} e r_\text{{LR}}}\mathcal{P}_\text{LR}^{(m)}(e) \, , 
\end{align}
where $\mu$ is the mass of the disc particle and $e$ is the (small) eccentricity of the test particle in the disc. 
The power amplitude $\mathcal{P}_\text{LR}^{(m)}$ characterises the cycle-averaged work that the metric perturbation from the secondary would do on a test particle on an orbit with a small eccentricity, $e$, when the epicyclic frequency is commensurate with the secondary’s perturbation frequency.
In the limit of small $e$, the power amplitude is linear in $e$, and the resonant amplitudes are independent of the actual value of $e$.

Hirata demonstrated that this power amplitude can be associated with the power dissipated to the future horizon and future null infinity through the constructive interference of GWs, as expressed in terms of Teukolsky amplitudes (which is different from the Lindblad resonance power). The Teukolsky amplitudes producing the constructive interference are the ones sourced by the secondary at $r=r_s$ in circular motion (with $n = 0$) and a particle in the disc on a slightly eccentric orbit (with mode number $n=\pm 1$) at $r = r_\text{LR}$, where the $+$ ($-$) corresponds to outer (inner) Lindblad resonances. 

For inner Lindblad resonances, the power amplitude is the negative of the power due to constructive interference emitted to infinity and down the horizon; i.e.,
\begin{align}
\sum_m \mathcal{P}^{(m)}_\text{ILR} &= - \delta P_\infty - \delta P_\text{H} \, ,  \\
\delta P_\infty &= \text{Re} \sum_{\ell m} \frac{c_{13}Z^\mathrm{H}_{\ell m 0} (Z^\infty_{\ell m -1})^*}{(m\Omega_s)^2} \,, \\
\delta P_\text{H} &= \text{Re} \sum_{\ell m} \frac{\alpha Z^\mathrm{H}_{\ell m 0} (Z^\mathrm{H}_{\ell m -1})^*}{(m\Omega_s)^2} \, .
\end{align}
Here, the $n =0$ solutions are evaluated with the secondary as source term, while the $n=-1$ solutions are evaluated with the test particle in a slightly eccentric orbit as source term. Using symmetry arguments between coefficients and amplitudes with $\pm m$, the power associated with inner Lindblad resonances can be written as
\begin{align}\label{eq:P_ILR}
\mathcal{P}^{(m)}_\text{ILR} = 
    -\frac{1}{(m \Omega_s)^2} \sum_{\ell \geq m}^{\infty} Z^{\mathrm{H}}_{\ell m 0} 
    \left(
        c_{13} Z^{\mathrm{\infty}*}_{\ell m -1} + \alpha Z^{\mathrm{H}*}_{\ell m -1}
    \right)
.
\end{align}
Following a similar procedure, the power associated with the outer Lindblad resonances is 
\begin{align}\label{eq:P_OLR}
\mathcal{P}^{(m)}_\text{OLR} = 
    -\frac{1}{(m \Omega_s)^2} \sum_{\ell \geq m}^{\infty} Z^{\mathrm{\infty}}_{\ell m 0} 
    \left(
        Z^{\mathrm{\infty}*}_{\ell m 1} - c^{*}_{13} Z^{\mathrm{H}*}_{\ell m 1}
    \right)
.
\end{align}

In \cref{eq:torqueSm}, Hirata invoked a subtle energy balance law relating the power from constructive interference to the change in energy and angular momentum of the secondary object. One key aspect of this balance law is that the energy (and angular momentum) in the disc is invariant under the Lindblad resonance at leading order. This follows from Hirata's assumptions that the disc particles follow circular geodesics of the Kerr spacetime, and the Lindblad resonances only perturb the orbit to linear order in eccentricity. Because, to linear order in $e$, circular and slightly eccentric orbits in Kerr have the same energy and angular momentum, Lindblad resonances do not change the energy and angular momentum of the disc under Hirata's assumptions.

Note that the power amplitude is not itself the power dissipated by a Lindblad resonance---that quantity arises only at second order, after the disc develops an eccentric response. This distinction explains why the torque strength, \cref{eq:TLRm}, scales as 
$|\mathcal{P}_\text{LR}^{(m)}|^2$: the first power amplitude factor comes from the perturbing force, while the second comes from the disc particle’s radial displacement, which is itself proportional to the forcing amplitude. In this sense, Hirata’s “power amplitude” should be viewed as a linear coupling coefficient, whose square determines the true resonant torque.

\begin{figure*}[ht]
\includegraphics[width=0.495\linewidth]{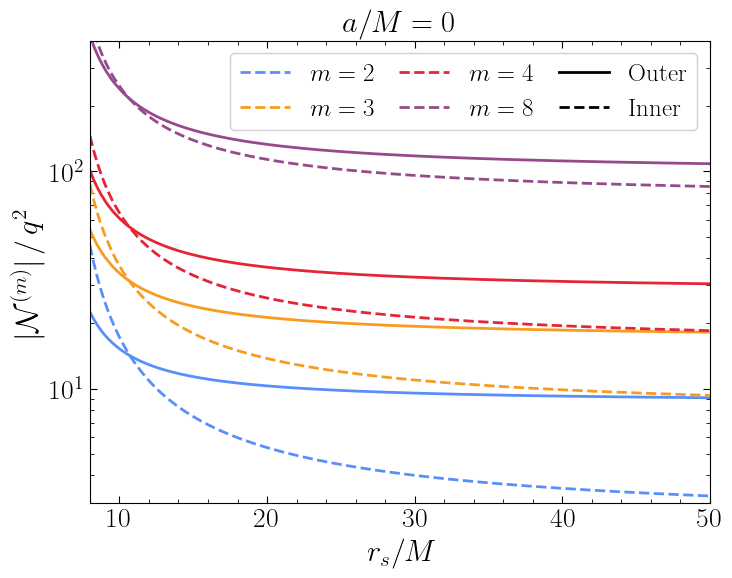}   
\includegraphics[width=0.495\linewidth]{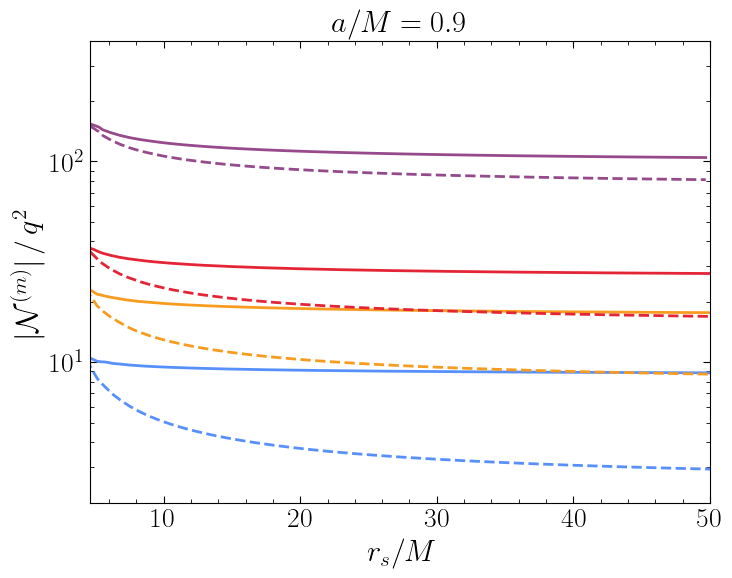}    
\caption{Normalised Lindblad torques as a function of the EMRI radius for different modes $m$. For large $r_s\gg r_+$, $\mathcal{N}^{(m)}$ approaches its constant Newtonian limit. For smaller $r_s$, the inner resonances become stronger than the outer ones, as anticipated from their location in Fig.~\ref{fig:LRlocations}. For large $m$, the amplitudes of the inner and outer torques become comparable, as the resonances tend to become equidistant from the secondary. Increasing primary spin leads to smaller normalised Lindblad torques.}
\label{fig:NormalisedTorqueStrengths}
\end{figure*}

We now have all the necessary ingredients to calculate the variable of interest, the torque strength, \cref{eq:TLRm}. It is often useful to calculate the normalised Lindblad resonance torque strength, $\mathcal{N}^{(m)}$, where the disc density is normalised out. This is given by~\cite{Hirata:2010vp}
\begin{align}
\mathcal{N}^{(m)} = \frac{T_\text{LR}^{(m)}}{2 \pi   r_\text{LR} \Sigma(r_\text{LR} ) } = \pm \frac{\pi m w^t \mathcal{Z}_\text{LR}}{|{D}'(r_{\text{LR}})|} \left|\mathcal{S}^{(m)} \right|^2 \, , \label{eq:NormalisedLindbladTorques}
\end{align}
which is independent of $r_s$ in the Newtonian limit, $r_s\gg r_+$.

\subsubsection{Results}

With the theoretical formalism established, the computation of Lindblad torques requires only a numerical Teukolsky solver. We employ the publicly available \texttt{pybhpt} Python package \cite{Nasipak:2025tby}, which provides all the necessary amplitudes and coefficients to evaluate Eq.~\eqref{eq:NormalisedLindbladTorques}. We verified that our results recover those presented in Table~2 of Hirata II~\cite{Hirata:2010vp}. In Eqs.~\eqref{eq:P_ILR} and \eqref{eq:P_OLR}, we typically sum up to $\ell_\text{max}=10(m - 1)$ to obtain converging results.

In Fig.~\ref{fig:NormalisedTorqueStrengths}, we plot the normalised torque strengths as a function of the secondary orbital radius for different values of $m$ and primary spin $a$. As noted in Hirata's original work, relativistic effects enhance the torque strength by a factor of a few with respect to the Newtonian limit (i.e., the constant value that $\mathcal{N}$ approaches at large $r$). Also, as anticipated in Sec.~\ref{sec:ResLocation}, the inner Lindblad torques become stronger than the outer ones at small $r_s$, as the location of the inner resonance moves closer to the secondary. However, note that this switch in torque strengths does not occur exactly at the switch in the resonant radius
, which we attribute to relativistic effects in the torque amplitude. Also, from Eq.~\eqref{eq:NormalisedLindbladTorques} we expect that the additional radial dependence in the actual torque $T_\text{LR}$ (e.g., through the disc density) will also affect the switch radius. The difference between inner and outer torque strengths becomes smaller for larger $m$ because, as mentioned, resonances tend to become equidistant from the secondary. 

From the right panel of Fig.~\ref{fig:NormalisedTorqueStrengths} we see that, on a given (prograde) orbit, the primary spin reduces the relativistic enhancement of the torque strengths. Note, however, that torque strengths will be further enhanced by the primary spin when the orbit is retrograde, see \cite{Hirata:2010vp}. As also anticipated from the resonance locations in Fig.~\ref{fig:LRlocations}, the switch between outer and inner torques happens at smaller radii for rapidly spinning primaries. 

\begin{figure*}[t]
\includegraphics[width=0.495\linewidth]{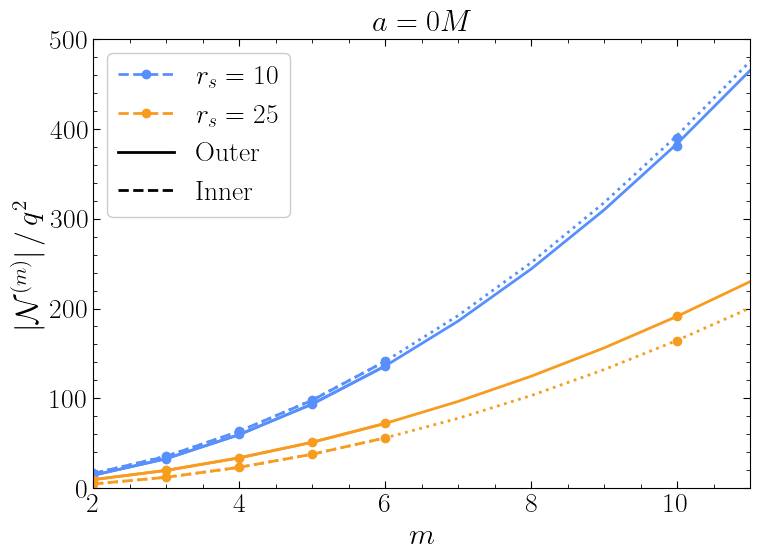}  
\includegraphics[width=0.495\linewidth]{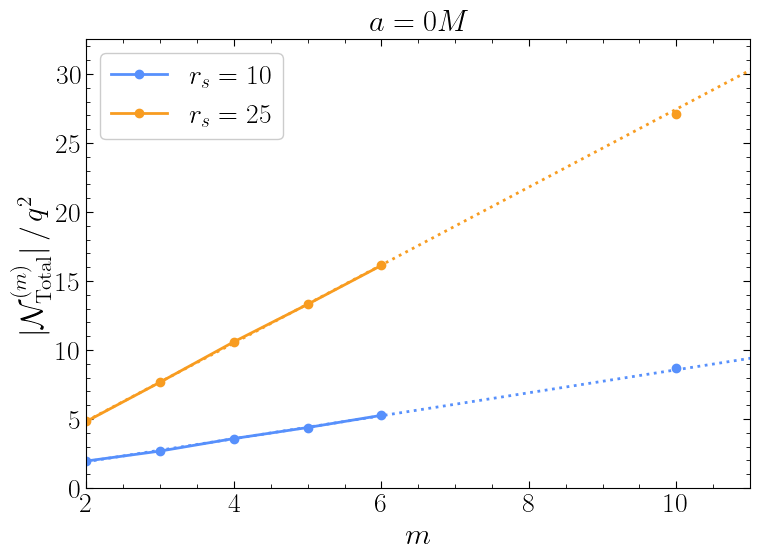}  
\caption{\textit{Left panel:} Variation of the mode-by-mode inner and outer normalised Lindblad torques with $m$. The growth becomes quadratic for any secondary radius $r_s$ and spin $a$ at relatively small $m$. \textit{Right panel:} same as in the left panel but for the total torque, i.e., the sum of the inner and outer contributions. In this case, the growth is approximately linear with $m$ because the quadratic components of the outer and inner torques cancel each other.}
\label{fig:TorqueGrowthWithm}
\end{figure*}

Since the larger the $m$, the closer the resonant radius is to the secondary, the amplitude of the torques grows with $m$. In Fig.~\ref{fig:TorqueGrowthWithm}, we present this growth for a representative orbital radius of the companion and primary BH spin. We find that the growth of the inner and outer Lindblad torques with $m$ for fixed secondary radius is quadratic already at small $m$, and the total normalised Lindblad torque (i.e., their sum) is approximately linear (Refs.~\cite{HegadeKR:2025dur, HegadeKR:2025rpr} presents an analytical derivation of this scaling in the large $m$ limit). We verified this happens for any spin and even at larger orbital radii. 
This quadratic scaling provides a substantial computational advantage in evaluating the total Lindblad torque at a given radius (summed over $m$). Rather than computing high-$m$ modes, which require prohibitively many $\ell$-modes, it suffices to calculate the torques at small $m$, perform a quadratic fit, and extrapolate to larger $m$. Without this simplification, obtaining the total torques would be numerically infeasible, as solving the Teukolsky equation at large $\ell$ is challenging.  

It is evident that, even in the Newtonian limit, the sum over $m$ for the total Lindblad torque diverges as $m \rightarrow \infty$. This divergence, well known in the migration literature, is non-physical and is regularized when pressure effects are considered, as we discuss in the next section.

%

\section{Relativistic Lindblad torques with Pressure corrections}\label{sec:TorquePressure}

\subsection{Pressure cut-off}
Astrophysical thin accretion discs are not strictly two-dimensional but possess a vertical structure characterised by the scale height $H$ (or, equivalently, the aspect ratio $h = H/r$). This thickness is maintained by pressure support, which also governs the propagation of density waves throughout the disc. Lindblad resonances excite such density waves.  

When summing over different Lindblad resonances at varying $m$, a key assumption is that their associated wavelengths and positions remain sufficiently distinct to avoid interference. Intuitively, this approximation fails once the wavelength of the perturbation becomes comparable to, or shorter than, the spacing between consecutive Lindblad resonances, leading to destructive interference and a weakened effective coupling between the perturber and the disc. In addition, pressure alters the angular velocity of disc particles, whose motion is no longer purely geodesic (or Keplerian in the Newtonian limit).  

In the Newtonian regime, accounting for pressure modifies the resonance detuning function to \cite{Artymowicz1993,Ward1997}  
\begin{align}
D_\text{Newt.}^\text{pressure}[r] &= m(\Omega[r] - \Omega_s)\mp \kappa_\text{Newt.}[r]\sqrt{1+\xi[r]^2}\, ,
\end{align}  
where the dimensionless pressure parameter is  
\begin{align}
\xi = m \frac{c_s}{\Omega\, r} = m \frac{H}{r} = m h \, ,
\end{align}  
and $c_s$ is the local sound speed. In the second equality, we used the fact that for a locally isothermal disc in vertical hydrostatic equilibrium, $c_s = H\Omega$.  

As a result of this modification, Lindblad resonances no longer asymptotically converge to the secondary’s orbital radius as $m \to \infty$. Instead, they saturate at a finite separation set by the disc scale height~\cite{Artymowicz1993}
\begin{equation}
r_\text{LR} \to r_s \pm \frac{2H}{3} \, .  
\end{equation}  
Relativistic effects are expected to slightly shift these saturation distances, but the correction should remain small since the effect is local to the secondary. For reference, thin accretion discs typically have aspect ratios $h \sim 0.01$--$0.1$~\cite{Gangardt:2024bic}.  

These pressure-induced modifications were first systematically studied in the seminal work of Goldreich and Tremaine on disc--planet interactions~\cite{GoldreichTremaine1979A}, and later refined by Artymowicz~\cite{Artymowicz1993}. In practice, torque amplitudes for modes satisfying  
\begin{align}
\xi \gtrsim 1 \;\;\Leftrightarrow\;\; m \gtrsim 1/h \sim 10\text{--}100 \, ,
\end{align}  
are strongly suppressed, due both to the displacement of the resonance locations and to pressure effects on the excitation amplitudes themselves. 

This suppression produces a cut-off function for the torque that combines polynomial and exponential decay. In particular, we adopt the following mode-by-mode prescription for the pressure cutoff\footnote{Alternative, one can introduce a cut-off at $\xi=1$, as done in Refs.~\cite{HegadeKR:2025dur, HegadeKR:2025rpr}.}
\begin{align}
T^{(m)}_\text{LR, pressure} \;=\; \left(\frac{2}{1 + e^\xi}\right)\left(\frac{1}{1 + 4\xi^2}\right) T^{(m)}_\text{LR} \, .\label{eq:TPressureCutoff}
\end{align}  
The numerical coefficients here may be tuned to match hydrodynamical simulations, though variations are expected to be only of order unity. With this decay, the total Lindblad torque summed over modes converges, enabling the use of Hirata’s formalism to assess the implications of relativistic Lindblad torques for GW astronomy.

\subsection{Disc models}

To compute the total torque, we need to specify a model for the disc density and scale height profiles. In this work, we use the relativistic model for a thin disc by Novikov and Thorne \cite{Novikov:1973kta}. For comparison, we also employ the non-relativistic Shakura-Sunyaev model \cite{Shakura:1972te}. 


%
\begin{figure*}[ht]
\includegraphics[width=0.495\linewidth]{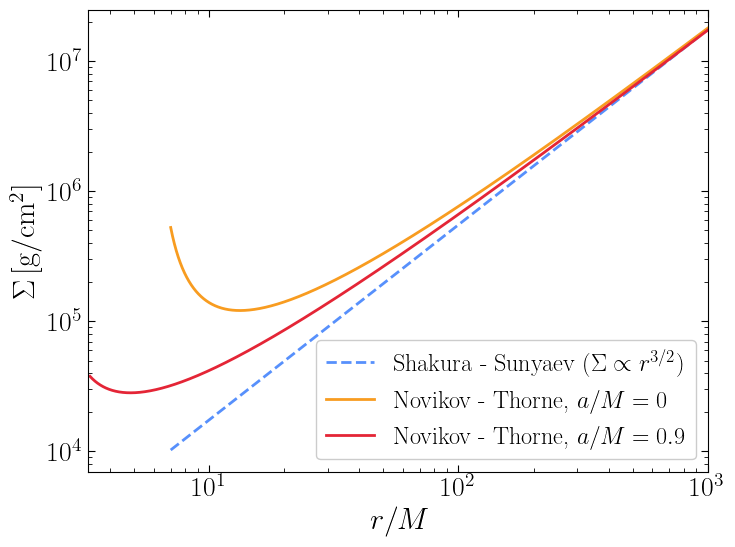}    
\includegraphics[width=0.495\linewidth]{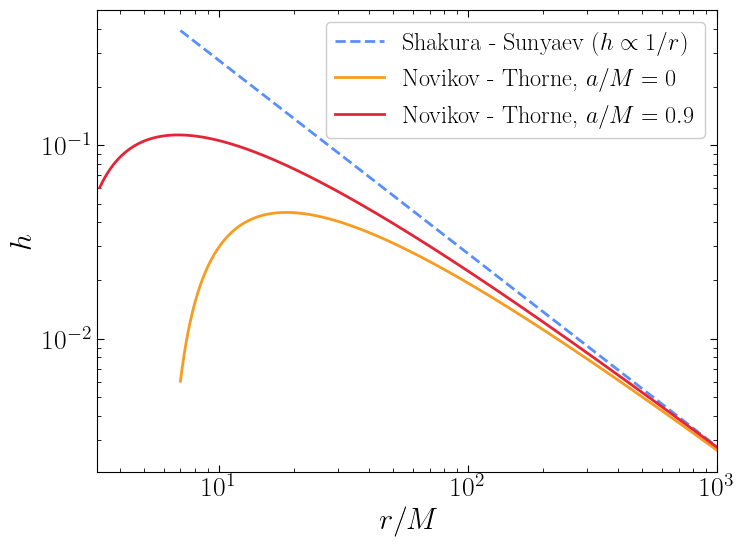}    
\caption{Surface density (left) and aspect ratio (right) radial profiles for the Novikov--Thorne disc model at different BH spins. The physical parameters of the system are $M = 10^6 \, M_\odot$ for the primary mass, $\dot{M}/\dot{M}_\text{Edd}= 0.1$ for the accretion rate in terms of the Eddington limit, and $\alpha = 0.1$ for the phenomenological parameter controlling viscosity. Changing these values alters the overall normalisation of the profiles but not their radial dependence. Dashed blue curves show the corresponding Shakura--Sunyaev power law profiles. Profiles are plotted to large radius, beyond the inner disc region---where radiation pressure dominates--- illustrating that relativistic effects are noticeable already for $r \lesssim 250M$.}
\label{fig:discProfile}
\end{figure*}

\paragraph{Shakura--Sunyaev.} 
The Shakura--Sunyaev $\alpha$-disc describes a geometrically thin, optically thick, radiatively efficient disc with axisymmetric, stationary flow on Keplerian circular orbits. In this regime, the surface density $\Sigma$ and the disc scale height (or aspect ratio) $h \equiv H/r$ can be represented by simple power laws
\begin{align}
\Sigma(r) &= \Sigma_0 \left( \frac{r}{10M} \right)^{-\Sigma_p}, \label{eq:DiscSSSigma} \\
h(r) &= h_0 \left( \frac{r}{10M} \right)^{(2\Sigma_p - 1)/4}. \label{eq:DiscSSh}
\end{align}
The exponent $\Sigma_p$ is set by the dominant pressure and opacity sources, while the normalizations $\Sigma_0$ and $h_0$ are fixed by global properties such as the accretion rate and viscosity. The aspect-ratio scaling is chosen so that the accretion rate is radially constant in steady state. In the inner disc ($r \lesssim 10^2\,M$), where radiation pressure dominates and electron scattering sets the opacity, one has $\Sigma_p = -3/2$, implying $\Sigma \propto r^{3/2}$ and $h \propto r^{-1}$.

\paragraph{Novikov--Thorne.} 
The Novikov--Thorne model is the general-relativistic extension of the Shakura--Sunyaev thin disc, incorporating orbital dynamics in the Kerr metric, including frame dragging, and adopting the ISCO as the inner boundary of the thin disc. In the standard prescription, the viscous torque is set to vanish at the ISCO, motivated by the rapid inward plunge; in practice, fluid elements can complete several orbits inside the ISCO, yielding a small but nonzero stress in the plunging region. For discs without a dynamically important net magnetic field, the Novikov-Thorne solution accurately describes the radial profiles outside the ISCO, in particular for $r \gtrsim r_{\rm ISCO} + M$~\cite{Potter:2021vlg}.

The Novikov-Thorne inner-disc profiles are given explicitly in Sec.~5.3 of Ref.~\cite{Abramowicz:2011xu} (see Eqs.~(99) therein). We plot their radial behaviour in Fig.~\ref{fig:discProfile} for two values of the central BH spin. We use as primary mass $M = 10^6 M_\odot$, accretion rate in terms of the Eddington ratio $\dot{M}/\dot{M}_\text{Edd}= 0.1$ and the viscosity parameter $\alpha = 0.1$. These choices are consistent with current observational constraints~\cite{Garcia:2016wse}, which, however, only probe the surface/atmosphere density of AGN discs and not the midplane where the secondary is located. Changing these values alters the overall normalisation of the profiles but not their radial dependence. We extend the plot to very large radii, to normalize all profiles to the same value in the Newtonian limit as $r \to \infty$. For comparison, we overlay the Shakura--Sunyaev power-law. 

The figure shows that, across the radii relevant for EMRIs in the LISA band, relativistic corrections to the disc structure are already significant, and the Novikov-Thorne profiles should be used in place of purely Newtonian scalings when studying the effects of migration torques. In particular, the density profiles are not captured by simple power laws and thus neither should migration torques, as they occur in Newtonian theory. We also stress that these are steady-state profiles, but AGN discs are expected to exhibit periods of turbulent flows, which can severely alter their stationary matter distribution configuration and impact the torques derived within the linear theory~\cite{Wu:2023qeh}.

\subsection{Results on total torques}

%
\begin{figure*}[t]
\includegraphics[width=0.495\linewidth]{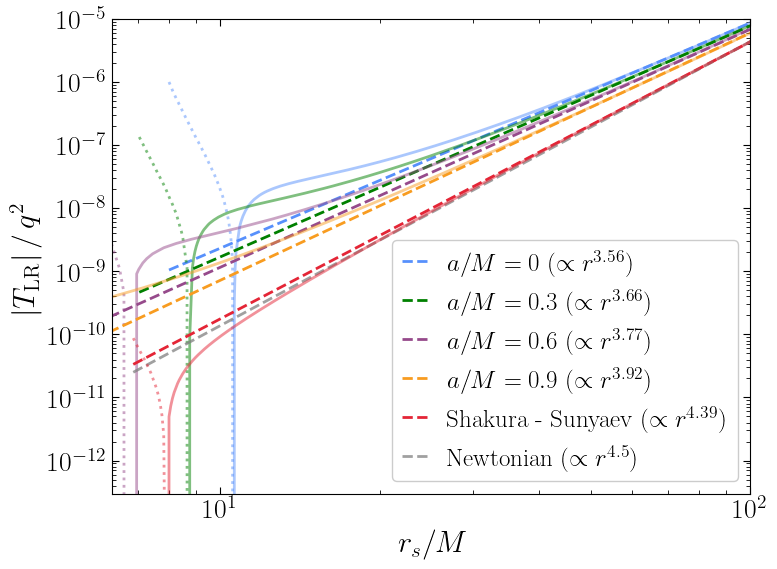}   
\includegraphics[width=0.495\linewidth]{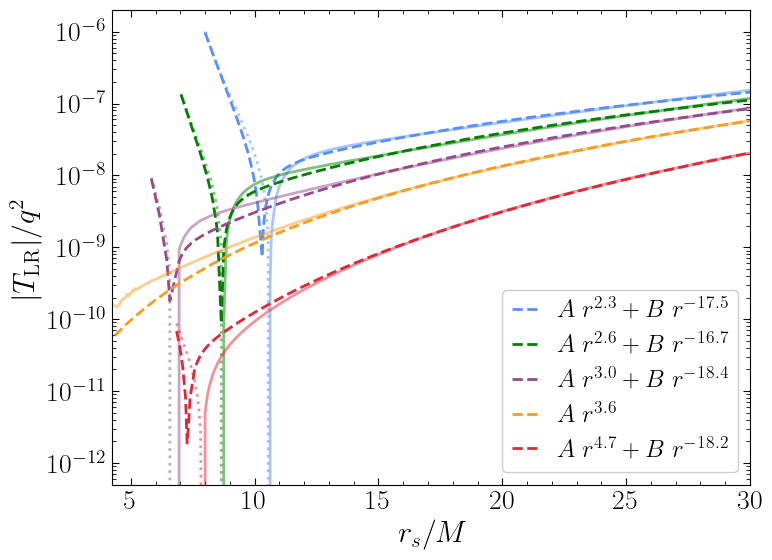}   
\caption{\textit{Left panel}: Relativistic Lindblad torque exerted on an equatorial, circular EMRI with radius $r_s$ embedded in different accretion disc profiles. Distinct spin values correspond to different Novikov–Thorne discs, normalised as in Fig.~\ref{fig:discProfile} (the same normalisation is applied to the Shakura–Sunyaev profile). Dotted curves denote opposite sign values of $T_\text{LR}$, where inner resonances dominate over outer ones, driving outward migration (though the EMRI still inspirals, since GW emission dominates its evolution). Dashed lines show power-law fits of the form $T_\text{LR} = A (r/10M)^{n^A_r}$. Same as in the left panel, but restricted to numerical data in the strong-field regime ($r_s < 30M$). Here, the Lindblad torque is fitted with a superposition of two power laws, $T_\text{LR} = A (r/10M)^{n^A_r} + B  (r/10M)^{n^B_r}$, to attempt capturing the transition between positive and negative $T_\text{LR}$. }
\label{fig:TorqueStrengths}
\end{figure*}

We compute the Lindblad torque acting on the EMRI using the density and aspect-ratio profiles shown in Fig.~\ref{fig:discProfile}. The torque is obtained by evaluating Eq.~\eqref{eq:TPressureCutoff} and summing over all azimuthal modes required for the pressure-corrected expression to converge (typically up to $m_\text{max} \sim 3/h$).

The results are displayed in Fig.~\ref{fig:TorqueStrengths}. A key feature, also highlighted in the recent works~\cite{HegadeKR:2025dur, HegadeKR:2025rpr}, is that relativistic torques in relativistic discs can differ by orders of magnitude from their Newtonian counterparts. In our Fig.~\ref{fig:TorqueStrengths}, part of this variation arises from the normalisation of the disc profiles in Fig.~\ref{fig:discProfile}: while these profiles coincide in the Newtonian limit ($r \to \infty$), they differ substantially in the strong-field regime. This discrepancy is not only due to differences in surface densities but also in scale heights, which determine the pressure cut-off and thus the number of contributing Lindblad resonances. Choosing an alternative normalisation, e.g.,~matching both surface densities and aspect ratios at a smaller radius (such as the ISCO, $r=10M$), still yields large differences in torque amplitudes at other radii. 

The radial dependence of the torque within each configuration is also noteworthy. As anticipated in the previous section, the Lindblad torque can change sign when the inner Lindblad contribution dominates over the outer one. In this regime, the net migration torque drives outward motion (as also noted in Ref.~\cite{HegadeKR:2025dur}). For the disc models considered here, however, GW emission remains the dominant driver of evolution (see Fig.~\ref{fig:TorqueGWComp}): although Lindblad torques act outward, the binary will still inspiral, albeit more slowly than in vacuum. Near the ISCO, the inner Lindblad torque diverges and could, in principle, exceed GW losses, but our disc and torque prescriptions are not valid at such radii. We therefore restrict our results to $r > r_\text{ISCO}+2M$. The radius at which the torque changes sign depends on the disc profile, and in Novikov--Thorne discs, it also depends on the primary spin. Increasing spin shifts the inversion radius inward towards the ISCO; for $a/M=0.9$ no inversion occurs within the range of radii shown (though one may appear at smaller separations).

We also explored whether Lindblad torques can be approximated by power laws of the form $T_\text{LR}=A(r_s/10M)^{n_r}$. Such parameterizations are valuable: given the wide range of possible physical models for environmental effects, it is unlikely that all physically motivated prescriptions can be implemented in LISA data analysis pipelines. Power-law fits provide a practical, model-agnostic description that can later be mapped onto more detailed models~\cite{Speri:2022upm}. In Newtonian theory \cite{Tanaka_2002,Tanaka3}
\begin{align}
    T^\text{Newt}_\text{LR} = c_1 M \Sigma \frac{r}{h^2} \, ,
\end{align}
with $c_1$ a constant calibrated to simulations and dependent on density and height gradients. For the Shakura--Sunyaev profile~\eqref{eq:DiscSSSigma}--\eqref{eq:DiscSSh}, the inner-disc scaling ($\Sigma_p=-3/2$) yields $T^\text{Newt.}_\text{LR} \propto r^{9/2}$.

To compare with this prediction, we fit our numerical torques to power laws in two regimes: (i) all radii with $r_s<100M$ (left panel of Fig.~\ref{fig:TorqueStrengths}), and (ii) the strong-field region $r_s<30M$ (right panel), where we used a sum of two power-laws, $T_\text{LR}=A(r_s/10M)^{n^A_r}+B(r_s/10M)^{n^B_r}$, to capture the transition across the torque sign change. Fit parameters are listed in Table~\ref{tab:FitParameters}.

For the Shakura--Sunyaev disc, neither fit reproduces the Newtonian index $9/2$ exactly, but the deviation is only at the $\sim 1\%$ level. In the strong-field case, the two-power-law model does not perfectly capture the transition radius, though less strongly than for Novikov--Thorne profiles. For the latter, fitting over $r_s<100M$ can underestimate the strong-field torque by nearly an order of magnitude, especially at low spin, with power-law indices differing by $\sim 10$–$50\%$ depending on fit domain. The fitted indices also vary significantly with spin and are consistently smaller than the Newtonian value. Restricting to the strong-field region further reduces the index. For example, in Novikov--Thorne discs we obtain $T_\text{LR}\propto r^{3.92}$ ($r^{3.56}$) for $a/M=0.9$ ($a/M=0$) when fitting all radii, and $T_\text{LR}\propto r^{3.96}$ ($r^{2.3}$) in the strong-field range.

Finally, we explored different values of viscosity and accretion rate, which set the overall normalisation of the surface density and scale height. We found that modifying the relation between $\Sigma$ and $h$ does not affect the rising power-law dependence of the torques with $r$, but it can slightly shift the radius at which the torque changes sign. This shift, however, is only of order $\mathcal{O}(M)$. 


\subsection{Consequences for GW astronomy}

\begin{table}[t]
\centering
\begin{tabular}{c|c|c|c|c|c}

$a/M$ & Fit domain & $A [10^{-9}]$ & $n^A_r$ & $-B [10^{-9}]$ & $-n^B_r$ \\ \hline

\multirow{2}{*}{$0$} 
& All $r_s$ & $ 2.33 $ & $3.57$ & -- & -- \\ 
& $r_s \leq 30M$ & 11.4 & 2.30 & 19.5 & 17.5 \\ \hline

\multirow{2}{*}{$0.3$} 
& All $r_s$ & $ 1.69 $ & $3.66$ & -- & -- \\ 
& $r_s \leq 30M$ & 6.35 & 2.61 &  0.383 & 16.7 \\ \hline

\multirow{2}{*}{$0.6$} 
& All $r_s$ & $ 1.16 $ & $3.77$ & -- & -- \\ 
& $r_s \leq 30M$ & 3.06 & 3.02 & $4.59 \text{E} -4$ & 18.4\\ \hline

\multirow{2}{*}{$0.9$} 
& All $r_s$ & $ 0.721$ & $3.92$ & -- & -- \\ 
& $r_s \leq 30M$ & 1.13 & 3.56 & -- & -- \\ \hline

\multirow{2}{*}{Shak-Suny.} 
& All $r_s$ & $0.176$ & $4.39$ & -- & -- \\ 
& $r_s \leq 30M$ & 0.124 & 4.65 & $9.24 \text{E} -5$ & 18.17\\ 

\end{tabular}
\caption{Coefficients of power-law fits $T_\text{LR} = A \,(r/10M)^{n_r^A}$ (or $T_\text{LR} = A \,(r/10M)^{n_r^A} + B \,(r/10M)^{n_r^B}  $) to the numerical data for Lindblad torques shown in Fig.~\ref{fig:TorqueStrengths}. Shak-Suny. denotes the Shakura-Sunyaev model.}
\label{tab:FitParameters}
\end{table}

The natural follow-up question to the previous section is how the relativistic enhancement of Lindblad torques in accretion discs impacts LISA observations of EMRIs. Building on earlier work, some qualitative conclusions can already be drawn.

Using state-of-the-art circular, equatorial EMRI waveforms in a fully Bayesian LISA data analysis, Ref.~\cite{Khalvati:2024tzz} showed that for a ``golden'' EMRI with $M = 10^6 M_\odot$, $\mu = 50 M_\odot$, $a/M = 0.9$, initial separation $r_0 \approx 15.5M$, and four years of observation before merger (at a sky location yielding $\text{SNR} = 50$), Newtonian Lindblad torques modelled as $T_\text{LR} = A \, r^{n_r}$ could be constrained with relative uncertainties of $\sim 15\%$ on the amplitude $A$ and $\sim 6\%$ on the slope $n_r$ for typical astrophysical disc configurations. Since relativistic corrections systematically enhance Lindblad torques relative to GW emission, we expect these constraints to improve, allowing LISA to probe a broader range of disc models. 

Importantly, the relative error on the slope $n_r$ is comparable to the differences found in our strong-field power-law fits (right panel of Fig.~\ref{fig:TorqueStrengths}). Moreover, because the slope depends on the spin of the central BH---measurable with exquisite precision in EMRI observations (errors $\sim 10^{-5}$)---LISA should be able to distinguish between relativistic Lindblad torques arising from different disc configurations. More quantitative conclusions require repeating the Bayesian inference studies of Refs.~\cite{Speri:2022upm,Khalvati:2024tzz} with relativistic torques injected into the waveforms and recovered with power-law models. We leave this investigation to future work.

A key caveat is the sign reversal of the Lindblad torque as the EMRI inspirals. Ref.~\cite{Speri:2022upm} (albeit with less accurate waveforms than Ref.~\cite{Khalvati:2024tzz}) found that for Newtonian torques, the parameters $A$ and $n_r$ are strongly correlated. This arises because the environmental effect is largest early in the inspiral, so GW observations are most sensitive to the combination $T_\text{LR} \approx A r_0^{n_r}$. This assumption holds provided there is only one stage of the inspiral where the relative contribution of Lindblad torques peaks.
In the relativistic case, however, the situation is more subtle: the torque can change sign and increase in magnitude near the ISCO, so a single power-law may not suffice. To assess this, Fig.~\ref{fig:TorqueGWComp} compares relativistic Lindblad torques with GW torques (computed using \texttt{few}~\cite{Chapman-Bird:2025xtd}) as a function of orbital radius. For the golden EMRI described above, only in low-spin Novikov–Thorne discs ($a/M = 0$ and $a/M = 0.3$) do the negative Lindblad torques near plunge become comparable (in relative magnitude) to those at the start of the inspiral ($r_s \gtrsim 15M$). For higher spins, we expect the conclusions of Ref.~\cite{Speri:2022upm} to hold, while at low spin the torque evolution is more complex and may require inference with multiple power laws. Observations and theory both suggest that luminous AGNs are typically powered by rapidly spinning SMBHs ($a/M \gtrsim 0.7$), though low-spin configurations remain possible in scenarios dominated by chaotic accretion or recent SMBH mergers~\cite{Berti:2008af, King:2008au}.

\section{Discussion}

We showed that relativistic corrections to Lindblad torques exerted on EMRIs embedded in accretion discs---both through modifications to the binary–disc interaction and to the disc structure---can amplify the torque amplitude by several orders of magnitude compared to Newtonian predictions. Moreover, relativistic torques display a different radial dependence, which itself varies with the primary BH spin. Together with previous studies on the detectability of accretion disc effects with LISA, our findings suggest that EMRI observations could, in principle, accurately measure disc properties and distinguish between different disc configurations.

As one of the first investigations of relativistic Lindblad torques on EMRIs---together with~\cite{HegadeKR:2025dur, HegadeKR:2025rpr}, to which our model should be quantitatively compared in the future---our work has some limitations:

\textbf{Disc modelling}. The torque model we adopted, following Hirata`s work, assumes a pressureless disc. As we discussed, this assumption leads to a divergence in the total torque. To regularise it, we introduced a phenomenological cut-off function inspired by Newtonian studies. However, a more rigorous treatment of pressure effects in relativistic discs is needed for more accurate predictions of disc torques. One promising approach---currently under exploration~\cite{DysonToAppear}---is to adapt the methodology developed in Ref.~\cite{dyson2025environmental} for EMRIs in superradiant scalar clouds: first reconstruct the perturbed spacetime metric from solutions of the (sourced) Teukolsky equation, and then solve the relativistic hydrodynamics equations (e.g., the Euler equation) for the disc in this background. This would yield fluid perturbations that encode Lindblad torques self-consistently, and, at the same time, capture additional effects absent from our treatment, such as accretion and co-orbital resonant torques near the secondary, which can compete with Lindblad resonances. 

We also assumed a stationary background disc model, whereas real AGN discs are expected to be turbulent and time-dependent. Strong turbulence could, in principle, invalidate linear perturbation theory estimates; but, for milder fluctuations, one can still capture the dominant physical effects \cite{Wu:2023qeh}. In such cases, the main consequence would be small biases in the inferred environmental parameters, rather than the vacuum EMRI ones \cite{Copparoni:2025jhq}. Developing a better understanding of the regime of validity of linear estimates across the parameter space will be necessary for robust astrophysical inferences. Numerical simulation will play a key role in that.

\textbf{Orbital configurations}. Our analysis was restricted to prograde, equatorial, circular orbits. Disc interactions are expected to damp inclination and drive circularisation by the time the EMRI enters the LISA band. Nonetheless, some residual eccentricity may survive~\cite{Spieksma:2025wex, 2024MNRAS.528.4958W}. Newtonian studies have shown that eccentricity enhances the detectability of disc effects with LISA~\cite{Duque:2024mfw}. Extending the relativistic analysis to eccentric orbits would therefore be a natural next step.

\textbf{Implications for LISA data analysis}. Finally, a crucial direction for future work is to determine how to incorporate environmental effects such as those studied here into LISA data analysis pipelines. A systematic survey across the relevant parameter space and for a range of disc models is needed to assess whether parametrised, agnostic waveform models can effectively capture the true physical signals. If the answer is positive, one could then establish a mapping between the agnostic waveforms and the physical models. Moreover, LISA is expected to detect $\sim 10^7$ overlapping signals, which must be extracted simultaneously through a global fit. An open question is whether small deviations from vacuum GR, including environmental effects, must be included directly in the global fit, or whether they can instead be studied in post-processing using a catalogue of sources obtained from vacuum-only templates. Addressing this will require accurate and physically well-motivated environmental models. 

\section{Acknowledgements}

We thank Abhishek Hegade K. R., Nicolás Yunes, and Conor Dyson for fruitful discussions and for generously sharing their results prior to publication. We also thank Zachary Nasipak for his guidance in the use of the package \texttt{pybhpt}, which was essential for efficiently computing the relativistic Lindblad torques. We thank Diogo Ribeiro for valuable discussions on numerical implementations. We thank Thomas Spieksma and Conor Dyson for carefully reading the manuscript and providing valuable feedback.
RV gratefully acknowledges the support of the Dutch Research Council
(NWO) through an Open Competition Domain Science-M grant, project number OCENW.M.21.375.
AS acknowledges support from the STFC Consolidated Grant no. ST/V005596/1.
LS acknowledges support from the UKRI Horizon guarantee funding (project no. EP/Y023706/1). LS is also supported by a University of Nottingham Anne McLaren Fellowship. LS would like to thank the TAPIR group at Caltech for their kind hospitality during the initial stages of this work.

\bibliography{bib}

\begin{thebibliography}{83}%
\makeatletter
\providecommand \@ifxundefined [1]{%
 \@ifx{#1\undefined}
}%
\providecommand \@ifnum [1]{%
 \ifnum #1\expandafter \@firstoftwo
 \else \expandafter \@secondoftwo
 \fi
}%
\providecommand \@ifx [1]{%
 \ifx #1\expandafter \@firstoftwo
 \else \expandafter \@secondoftwo
 \fi
}%
\providecommand \natexlab [1]{#1}%
\providecommand \enquote  [1]{``#1''}%
\providecommand \bibnamefont  [1]{#1}%
\providecommand \bibfnamefont [1]{#1}%
\providecommand \citenamefont [1]{#1}%
\providecommand \href@noop [0]{\@secondoftwo}%
\providecommand \href [0]{\begingroup \@sanitize@url \@href}%
\providecommand \@href[1]{\@@startlink{#1}\@@href}%
\providecommand \@@href[1]{\endgroup#1\@@endlink}%
\providecommand \@sanitize@url [0]{\catcode `\\12\catcode `\$12\catcode
  `\&12\catcode `\#12\catcode `\^12\catcode `\_12\catcode `\%12\relax}%
\providecommand \@@startlink[1]{}%
\providecommand \@@endlink[0]{}%
\providecommand \url  [0]{\begingroup\@sanitize@url \@url }%
\providecommand \@url [1]{\endgroup\@href {#1}{\urlprefix }}%
\providecommand \urlprefix  [0]{URL }%
\providecommand \Eprint [0]{\href }%
\providecommand \doibase [0]{https://doi.org/}%
\providecommand \selectlanguage [0]{\@gobble}%
\providecommand \bibinfo  [0]{\@secondoftwo}%
\providecommand \bibfield  [0]{\@secondoftwo}%
\providecommand \translation [1]{[#1]}%
\providecommand \BibitemOpen [0]{}%
\providecommand \bibitemStop [0]{}%
\providecommand \bibitemNoStop [0]{.\EOS\space}%
\providecommand \EOS [0]{\spacefactor3000\relax}%
\providecommand \BibitemShut  [1]{\csname bibitem#1\endcsname}%
\let\auto@bib@innerbib\@empty
\bibitem [{\citenamefont {Neumayer}\ \emph {et~al.}(2020)\citenamefont
  {Neumayer}, \citenamefont {Seth},\ and\ \citenamefont
  {Boeker}}]{Neumayer:2020gno}%
  \BibitemOpen
  \bibfield  {author} {\bibinfo {author} {\bibfnamefont {N.}~\bibnamefont
  {Neumayer}}, \bibinfo {author} {\bibfnamefont {A.}~\bibnamefont {Seth}},\
  and\ \bibinfo {author} {\bibfnamefont {T.}~\bibnamefont {Boeker}},\
  }\bibfield  {title} {\bibinfo {title} {{Nuclear star clusters}},\ }\href
  {https://doi.org/10.1007/s00159-020-00125-0} {\bibfield  {journal} {\bibinfo
  {journal} {Astron. Astrophys. Rev.}\ }\textbf {\bibinfo {volume} {28}},\
  \bibinfo {pages} {4} (\bibinfo {year} {2020})},\ \Eprint
  {https://arxiv.org/abs/2001.03626} {arXiv:2001.03626 [astro-ph.GA]}
  \BibitemShut {NoStop}%
\bibitem [{\citenamefont {Miller}\ and\ \citenamefont
  {Lauburg}(2009)}]{Miller:2008yw}%
  \BibitemOpen
  \bibfield  {author} {\bibinfo {author} {\bibfnamefont {M.~C.}\ \bibnamefont
  {Miller}}\ and\ \bibinfo {author} {\bibfnamefont {V.~M.}\ \bibnamefont
  {Lauburg}},\ }\bibfield  {title} {\bibinfo {title} {{Mergers of Stellar-Mass
  Black Holes in Nuclear Star Clusters}},\ }\href
  {https://doi.org/10.1088/0004-637X/692/1/917} {\bibfield  {journal} {\bibinfo
   {journal} {Astrophys. J.}\ }\textbf {\bibinfo {volume} {692}},\ \bibinfo
  {pages} {917} (\bibinfo {year} {2009})},\ \Eprint
  {https://arxiv.org/abs/0804.2783} {arXiv:0804.2783 [astro-ph]} \BibitemShut
  {NoStop}%
\bibitem [{\citenamefont {Abramowicz}\ and\ \citenamefont
  {Fragile}(2013)}]{Abramowicz:2011xu}%
  \BibitemOpen
  \bibfield  {author} {\bibinfo {author} {\bibfnamefont {M.~A.}\ \bibnamefont
  {Abramowicz}}\ and\ \bibinfo {author} {\bibfnamefont {P.~C.}\ \bibnamefont
  {Fragile}},\ }\bibfield  {title} {\bibinfo {title} {{Foundations of Black
  Hole Accretion Disk Theory}},\ }\href {https://doi.org/10.12942/lrr-2013-1}
  {\bibfield  {journal} {\bibinfo  {journal} {Living Rev. Rel.}\ }\textbf
  {\bibinfo {volume} {16}},\ \bibinfo {pages} {1} (\bibinfo {year} {2013})},\
  \Eprint {https://arxiv.org/abs/1104.5499} {arXiv:1104.5499 [astro-ph.HE]}
  \BibitemShut {NoStop}%
\bibitem [{\citenamefont {Gondolo}\ and\ \citenamefont
  {Silk}(1999)}]{Gondolo:1999ef}%
  \BibitemOpen
  \bibfield  {author} {\bibinfo {author} {\bibfnamefont {P.}~\bibnamefont
  {Gondolo}}\ and\ \bibinfo {author} {\bibfnamefont {J.}~\bibnamefont {Silk}},\
  }\bibfield  {title} {\bibinfo {title} {{Dark matter annihilation at the
  galactic center}},\ }\href {https://doi.org/10.1103/PhysRevLett.83.1719}
  {\bibfield  {journal} {\bibinfo  {journal} {Phys. Rev. Lett.}\ }\textbf
  {\bibinfo {volume} {83}},\ \bibinfo {pages} {1719} (\bibinfo {year}
  {1999})},\ \Eprint {https://arxiv.org/abs/astro-ph/9906391}
  {arXiv:astro-ph/9906391} \BibitemShut {NoStop}%
\bibitem [{\citenamefont {Brito}\ \emph {et~al.}(2015)\citenamefont {Brito},
  \citenamefont {Cardoso},\ and\ \citenamefont {Pani}}]{Brito:2015oca}%
  \BibitemOpen
  \bibfield  {author} {\bibinfo {author} {\bibfnamefont {R.}~\bibnamefont
  {Brito}}, \bibinfo {author} {\bibfnamefont {V.}~\bibnamefont {Cardoso}},\
  and\ \bibinfo {author} {\bibfnamefont {P.}~\bibnamefont {Pani}},\ }\bibfield
  {title} {\bibinfo {title} {{Superradiance}: {New Frontiers in Black Hole
  Physics}},\ }\href {https://doi.org/10.1007/978-3-319-19000-6} {\bibfield
  {journal} {\bibinfo  {journal} {Lect. Notes Phys.}\ }\textbf {\bibinfo
  {volume} {906}},\ \bibinfo {pages} {pp.1} (\bibinfo {year} {2015})},\ \Eprint
  {https://arxiv.org/abs/1501.06570} {arXiv:1501.06570 [gr-qc]} \BibitemShut
  {NoStop}%
\bibitem [{\citenamefont {Hui}(2021)}]{Hui:2021tkt}%
  \BibitemOpen
  \bibfield  {author} {\bibinfo {author} {\bibfnamefont {L.}~\bibnamefont
  {Hui}},\ }\bibfield  {title} {\bibinfo {title} {{Wave Dark Matter}},\ }\href
  {https://doi.org/10.1146/annurev-astro-120920-010024} {\bibfield  {journal}
  {\bibinfo  {journal} {Ann. Rev. Astron. Astrophys.}\ }\textbf {\bibinfo
  {volume} {59}},\ \bibinfo {pages} {247} (\bibinfo {year} {2021})},\ \Eprint
  {https://arxiv.org/abs/2101.11735} {arXiv:2101.11735 [astro-ph.CO]}
  \BibitemShut {NoStop}%
\bibitem [{\citenamefont {Kocsis}\ \emph {et~al.}(2011)\citenamefont {Kocsis},
  \citenamefont {Yunes},\ and\ \citenamefont {Loeb}}]{Kocsis:2011dr}%
  \BibitemOpen
  \bibfield  {author} {\bibinfo {author} {\bibfnamefont {B.}~\bibnamefont
  {Kocsis}}, \bibinfo {author} {\bibfnamefont {N.}~\bibnamefont {Yunes}},\ and\
  \bibinfo {author} {\bibfnamefont {A.}~\bibnamefont {Loeb}},\ }\bibfield
  {title} {\bibinfo {title} {{Observable Signatures of EMRI Black Hole Binaries
  Embedded in Thin Accretion Disks}},\ }\href
  {https://doi.org/10.1103/PhysRevD.86.049907} {\bibfield  {journal} {\bibinfo
  {journal} {Phys. Rev. D}\ }\textbf {\bibinfo {volume} {84}},\ \bibinfo
  {pages} {024032} (\bibinfo {year} {2011})},\ \Eprint
  {https://arxiv.org/abs/1104.2322} {arXiv:1104.2322 [astro-ph.GA]}
  \BibitemShut {NoStop}%
\bibitem [{\citenamefont {Barausse}\ \emph {et~al.}(2015)\citenamefont
  {Barausse}, \citenamefont {Cardoso},\ and\ \citenamefont
  {Pani}}]{Barausse:2014pra}%
  \BibitemOpen
  \bibfield  {author} {\bibinfo {author} {\bibfnamefont {E.}~\bibnamefont
  {Barausse}}, \bibinfo {author} {\bibfnamefont {V.}~\bibnamefont {Cardoso}},\
  and\ \bibinfo {author} {\bibfnamefont {P.}~\bibnamefont {Pani}},\ }\bibfield
  {title} {\bibinfo {title} {{Environmental Effects for Gravitational-wave
  Astrophysics}},\ }\bibfield  {booktitle} {\emph {\bibinfo {booktitle}
  {{Proceedings, 10th International LISA Symposium}}},\ }\href
  {https://doi.org/10.1088/1742-6596/610/1/012044} {\bibfield  {journal}
  {\bibinfo  {journal} {J. Phys. Conf. Ser.}\ }\textbf {\bibinfo {volume}
  {610}},\ \bibinfo {pages} {012044} (\bibinfo {year} {2015})},\ \Eprint
  {https://arxiv.org/abs/1404.7140} {arXiv:1404.7140 [astro-ph.CO]}
  \BibitemShut {NoStop}%
\bibitem [{\citenamefont {Speri}\ \emph {et~al.}(2022)\citenamefont {Speri},
  \citenamefont {Antonelli}, \citenamefont {Sberna}, \citenamefont {Babak},
  \citenamefont {Barausse}, \citenamefont {Gair},\ and\ \citenamefont
  {Katz}}]{Speri:2022upm}%
  \BibitemOpen
  \bibfield  {author} {\bibinfo {author} {\bibfnamefont {L.}~\bibnamefont
  {Speri}}, \bibinfo {author} {\bibfnamefont {A.}~\bibnamefont {Antonelli}},
  \bibinfo {author} {\bibfnamefont {L.}~\bibnamefont {Sberna}}, \bibinfo
  {author} {\bibfnamefont {S.}~\bibnamefont {Babak}}, \bibinfo {author}
  {\bibfnamefont {E.}~\bibnamefont {Barausse}}, \bibinfo {author}
  {\bibfnamefont {J.~R.}\ \bibnamefont {Gair}},\ and\ \bibinfo {author}
  {\bibfnamefont {M.~L.}\ \bibnamefont {Katz}},\ }\bibfield  {title} {\bibinfo
  {title} {{Measuring accretion-disk effects with gravitational waves from
  extreme mass ratio inspirals}},\ }\href@noop {} {\  (\bibinfo {year}
  {2022})},\ \Eprint {https://arxiv.org/abs/2207.10086} {arXiv:2207.10086
  [gr-qc]} \BibitemShut {NoStop}%
\bibitem [{\citenamefont {Cole}\ \emph {et~al.}(2023)\citenamefont {Cole},
  \citenamefont {Bertone}, \citenamefont {Coogan}, \citenamefont {Gaggero},
  \citenamefont {Karydas}, \citenamefont {Kavanagh}, \citenamefont {Spieksma},\
  and\ \citenamefont {Tomaselli}}]{Cole:2022yzw}%
  \BibitemOpen
  \bibfield  {author} {\bibinfo {author} {\bibfnamefont {P.~S.}\ \bibnamefont
  {Cole}}, \bibinfo {author} {\bibfnamefont {G.}~\bibnamefont {Bertone}},
  \bibinfo {author} {\bibfnamefont {A.}~\bibnamefont {Coogan}}, \bibinfo
  {author} {\bibfnamefont {D.}~\bibnamefont {Gaggero}}, \bibinfo {author}
  {\bibfnamefont {T.}~\bibnamefont {Karydas}}, \bibinfo {author} {\bibfnamefont
  {B.~J.}\ \bibnamefont {Kavanagh}}, \bibinfo {author} {\bibfnamefont
  {T.~F.~M.}\ \bibnamefont {Spieksma}},\ and\ \bibinfo {author} {\bibfnamefont
  {G.~M.}\ \bibnamefont {Tomaselli}},\ }\bibfield  {title} {\bibinfo {title}
  {{Distinguishing environmental effects on binary black hole gravitational
  waveforms}},\ }\href {https://doi.org/10.1038/s41550-023-01990-2} {\bibfield
  {journal} {\bibinfo  {journal} {Nature Astron.}\ }\textbf {\bibinfo {volume}
  {7}},\ \bibinfo {pages} {943} (\bibinfo {year} {2023})},\ \Eprint
  {https://arxiv.org/abs/2211.01362} {arXiv:2211.01362 [gr-qc]} \BibitemShut
  {NoStop}%
\bibitem [{\citenamefont {Chen}\ \emph {et~al.}(2024)\citenamefont {Chen} \emph
  {et~al.}}]{NANOGrav:2024nmo}%
  \BibitemOpen
  \bibfield  {author} {\bibinfo {author} {\bibfnamefont {Y.}~\bibnamefont
  {Chen}} \emph {et~al.} (\bibinfo {collaboration} {NANOGrav}),\ }\bibfield
  {title} {\bibinfo {title} {{Galaxy Tomography with the Gravitational Wave
  Background from Supermassive Black Hole Binaries}},\ }\href@noop {} {\
  (\bibinfo {year} {2024})},\ \Eprint {https://arxiv.org/abs/2411.05906}
  {arXiv:2411.05906 [astro-ph.HE]} \BibitemShut {NoStop}%
\bibitem [{\citenamefont {Garg}\ \emph {et~al.}(2022)\citenamefont {Garg},
  \citenamefont {Derdzinski}, \citenamefont {Zwick}, \citenamefont {Capelo},\
  and\ \citenamefont {Mayer}}]{Garg:2022nko}%
  \BibitemOpen
  \bibfield  {author} {\bibinfo {author} {\bibfnamefont {M.}~\bibnamefont
  {Garg}}, \bibinfo {author} {\bibfnamefont {A.}~\bibnamefont {Derdzinski}},
  \bibinfo {author} {\bibfnamefont {L.}~\bibnamefont {Zwick}}, \bibinfo
  {author} {\bibfnamefont {P.~R.}\ \bibnamefont {Capelo}},\ and\ \bibinfo
  {author} {\bibfnamefont {L.}~\bibnamefont {Mayer}},\ }\bibfield  {title}
  {\bibinfo {title} {{The imprint of gas on gravitational waves from LISA
  intermediate-mass black hole binaries}}\ }\href
  {https://doi.org/10.1093/mnras/stac2711} {10.1093/mnras/stac2711} (\bibinfo
  {year} {2022}),\ \Eprint {https://arxiv.org/abs/2206.05292} {arXiv:2206.05292
  [astro-ph.GA]} \BibitemShut {NoStop}%
\bibitem [{\citenamefont {Caneva~Santoro}\ \emph {et~al.}(2024)\citenamefont
  {Caneva~Santoro}, \citenamefont {Roy}, \citenamefont {Vicente}, \citenamefont
  {Haney}, \citenamefont {Piccinni}, \citenamefont {Del~Pozzo},\ and\
  \citenamefont {Martinez}}]{CanevaSantoro:2023aol}%
  \BibitemOpen
  \bibfield  {author} {\bibinfo {author} {\bibfnamefont {G.}~\bibnamefont
  {Caneva~Santoro}}, \bibinfo {author} {\bibfnamefont {S.}~\bibnamefont {Roy}},
  \bibinfo {author} {\bibfnamefont {R.}~\bibnamefont {Vicente}}, \bibinfo
  {author} {\bibfnamefont {M.}~\bibnamefont {Haney}}, \bibinfo {author}
  {\bibfnamefont {O.~J.}\ \bibnamefont {Piccinni}}, \bibinfo {author}
  {\bibfnamefont {W.}~\bibnamefont {Del~Pozzo}},\ and\ \bibinfo {author}
  {\bibfnamefont {M.}~\bibnamefont {Martinez}},\ }\bibfield  {title} {\bibinfo
  {title} {{First Constraints on Compact Binary Environments from LIGO-Virgo
  Data}},\ }\href {https://doi.org/10.1103/PhysRevLett.132.251401} {\bibfield
  {journal} {\bibinfo  {journal} {Phys. Rev. Lett.}\ }\textbf {\bibinfo
  {volume} {132}},\ \bibinfo {pages} {251401} (\bibinfo {year} {2024})},\
  \Eprint {https://arxiv.org/abs/2309.05061} {arXiv:2309.05061 [gr-qc]}
  \BibitemShut {NoStop}%
\bibitem [{\citenamefont {Roy}\ and\ \citenamefont
  {Vicente}(2025)}]{Roy:2024rhe}%
  \BibitemOpen
  \bibfield  {author} {\bibinfo {author} {\bibfnamefont {S.}~\bibnamefont
  {Roy}}\ and\ \bibinfo {author} {\bibfnamefont {R.}~\bibnamefont {Vicente}},\
  }\bibfield  {title} {\bibinfo {title} {{Compact binary coalescences in dense
  gaseous environments can pose as ones in vacuum}},\ }\href
  {https://doi.org/10.1103/PhysRevD.111.084037} {\bibfield  {journal} {\bibinfo
   {journal} {Phys. Rev. D}\ }\textbf {\bibinfo {volume} {111}},\ \bibinfo
  {pages} {084037} (\bibinfo {year} {2025})},\ \Eprint
  {https://arxiv.org/abs/2410.16388} {arXiv:2410.16388 [gr-qc]} \BibitemShut
  {NoStop}%
\bibitem [{\citenamefont {Zwick}\ \emph {et~al.}(2025)\citenamefont {Zwick},
  \citenamefont {Tak{\'a}tsy}, \citenamefont {Saini}, \citenamefont {Hendriks},
  \citenamefont {Samsing}, \citenamefont {Tiede}, \citenamefont {Rowan},\ and\
  \citenamefont {Trani}}]{Zwick:2025wkt}%
  \BibitemOpen
  \bibfield  {author} {\bibinfo {author} {\bibfnamefont {L.}~\bibnamefont
  {Zwick}}, \bibinfo {author} {\bibfnamefont {J.}~\bibnamefont {Tak{\'a}tsy}},
  \bibinfo {author} {\bibfnamefont {P.}~\bibnamefont {Saini}}, \bibinfo
  {author} {\bibfnamefont {K.}~\bibnamefont {Hendriks}}, \bibinfo {author}
  {\bibfnamefont {J.}~\bibnamefont {Samsing}}, \bibinfo {author} {\bibfnamefont
  {C.}~\bibnamefont {Tiede}}, \bibinfo {author} {\bibfnamefont
  {C.}~\bibnamefont {Rowan}},\ and\ \bibinfo {author} {\bibfnamefont {A.~A.}\
  \bibnamefont {Trani}},\ }\bibfield  {title} {\bibinfo {title} {{Environmental
  effects in stellar mass gravitational wave sources I: Expected fraction of
  signals with significant dephasing in the dynamical and AGN channels}},\
  }\href@noop {} {\  (\bibinfo {year} {2025})},\ \Eprint
  {https://arxiv.org/abs/2503.24084} {arXiv:2503.24084 [astro-ph.HE]}
  \BibitemShut {NoStop}%
\bibitem [{\citenamefont {Tomaselli}(2025)}]{Tomaselli:2024ojz}%
  \BibitemOpen
  \bibfield  {author} {\bibinfo {author} {\bibfnamefont {G.~M.}\ \bibnamefont
  {Tomaselli}},\ }\bibfield  {title} {\bibinfo {title} {{Scattering of wave
  dark matter by supermassive black holes}},\ }\href
  {https://doi.org/10.1103/PhysRevD.111.063075} {\bibfield  {journal} {\bibinfo
   {journal} {Phys. Rev. D}\ }\textbf {\bibinfo {volume} {111}},\ \bibinfo
  {pages} {063075} (\bibinfo {year} {2025})},\ \Eprint
  {https://arxiv.org/abs/2501.00090} {arXiv:2501.00090 [gr-qc]} \BibitemShut
  {NoStop}%
\bibitem [{\citenamefont {Chen}\ \emph {et~al.}(2025)\citenamefont {Chen},
  \citenamefont {Chandramouli}, \citenamefont {Pozzoli}, \citenamefont
  {Buscicchio},\ and\ \citenamefont {Barausse}}]{Chen:2025qyj}%
  \BibitemOpen
  \bibfield  {author} {\bibinfo {author} {\bibfnamefont {R.}~\bibnamefont
  {Chen}}, \bibinfo {author} {\bibfnamefont {R.~S.}\ \bibnamefont
  {Chandramouli}}, \bibinfo {author} {\bibfnamefont {F.}~\bibnamefont
  {Pozzoli}}, \bibinfo {author} {\bibfnamefont {R.}~\bibnamefont
  {Buscicchio}},\ and\ \bibinfo {author} {\bibfnamefont {E.}~\bibnamefont
  {Barausse}},\ }\bibfield  {title} {\bibinfo {title} {{Muffled Murmurs:
  Environmental effects in the LISA stochastic signal from stellar-mass black
  hole binaries}},\ }\href@noop {} {\  (\bibinfo {year} {2025})},\ \Eprint
  {https://arxiv.org/abs/2507.00694} {arXiv:2507.00694 [gr-qc]} \BibitemShut
  {NoStop}%
\bibitem [{\citenamefont {Zwick}\ \emph {et~al.}(2021)\citenamefont {Zwick},
  \citenamefont {Derdzinski}, \citenamefont {Garg}, \citenamefont {Capelo},\
  and\ \citenamefont {Mayer}}]{Zwick:2021dlg}%
  \BibitemOpen
  \bibfield  {author} {\bibinfo {author} {\bibfnamefont {L.}~\bibnamefont
  {Zwick}}, \bibinfo {author} {\bibfnamefont {A.}~\bibnamefont {Derdzinski}},
  \bibinfo {author} {\bibfnamefont {M.}~\bibnamefont {Garg}}, \bibinfo {author}
  {\bibfnamefont {P.~R.}\ \bibnamefont {Capelo}},\ and\ \bibinfo {author}
  {\bibfnamefont {L.}~\bibnamefont {Mayer}},\ }\bibfield  {title} {\bibinfo
  {title} {{Dirty waveforms: multiband harmonic content of gas-embedded
  gravitational wave sources}},\ }\href@noop {} {\  (\bibinfo {year} {2021})},\
  \Eprint {https://arxiv.org/abs/2110.09097} {arXiv:2110.09097 [astro-ph.HE]}
  \BibitemShut {NoStop}%
\bibitem [{\citenamefont {Derdzinski}\ \emph {et~al.}(2021)\citenamefont
  {Derdzinski}, \citenamefont {D'Orazio}, \citenamefont {Duffell},
  \citenamefont {Haiman},\ and\ \citenamefont
  {MacFadyen}}]{Derdzinski:2020wlw}%
  \BibitemOpen
  \bibfield  {author} {\bibinfo {author} {\bibfnamefont {A.}~\bibnamefont
  {Derdzinski}}, \bibinfo {author} {\bibfnamefont {D.}~\bibnamefont
  {D'Orazio}}, \bibinfo {author} {\bibfnamefont {P.}~\bibnamefont {Duffell}},
  \bibinfo {author} {\bibfnamefont {Z.}~\bibnamefont {Haiman}},\ and\ \bibinfo
  {author} {\bibfnamefont {A.}~\bibnamefont {MacFadyen}},\ }\bibfield  {title}
  {\bibinfo {title} {{Evolution of gas disc\textendash{}embedded intermediate
  mass ratio inspirals in the $LISA$ band}},\ }\href
  {https://doi.org/10.1093/mnras/staa3976} {\bibfield  {journal} {\bibinfo
  {journal} {Mon. Not. Roy. Astron. Soc.}\ }\textbf {\bibinfo {volume} {501}},\
  \bibinfo {pages} {3540} (\bibinfo {year} {2021})},\ \Eprint
  {https://arxiv.org/abs/2005.11333} {arXiv:2005.11333 [astro-ph.HE]}
  \BibitemShut {NoStop}%
\bibitem [{\citenamefont {Santos}\ \emph {et~al.}(2025)\citenamefont {Santos},
  \citenamefont {Cardoso}, \citenamefont {Nat{\'a}rio},\ and\ \citenamefont
  {van~de Meent}}]{Santos:2025ass}%
  \BibitemOpen
  \bibfield  {author} {\bibinfo {author} {\bibfnamefont {J.~S.}\ \bibnamefont
  {Santos}}, \bibinfo {author} {\bibfnamefont {V.}~\bibnamefont {Cardoso}},
  \bibinfo {author} {\bibfnamefont {J.}~\bibnamefont {Nat{\'a}rio}},\ and\
  \bibinfo {author} {\bibfnamefont {M.}~\bibnamefont {van~de Meent}},\
  }\bibfield  {title} {\bibinfo {title} {{Gravitational waves from b-EMRIs:
  Doppler shift and beaming, resonant excitation, helicity oscillations and
  self-lensing}},\ }\href@noop {} {\  (\bibinfo {year} {2025})},\ \Eprint
  {https://arxiv.org/abs/2506.14868} {arXiv:2506.14868 [gr-qc]} \BibitemShut
  {NoStop}%
\bibitem [{\citenamefont {Hu}\ \emph {et~al.}(2023)\citenamefont {Hu},
  \citenamefont {Cai},\ and\ \citenamefont {Wang}}]{Hu:2023oiu}%
  \BibitemOpen
  \bibfield  {author} {\bibinfo {author} {\bibfnamefont {L.}~\bibnamefont
  {Hu}}, \bibinfo {author} {\bibfnamefont {R.-G.}\ \bibnamefont {Cai}},\ and\
  \bibinfo {author} {\bibfnamefont {S.-J.}\ \bibnamefont {Wang}},\ }\bibfield
  {title} {\bibinfo {title} {{Distinctive GWBs from eccentric inspiraling SMBH
  binaries with a DM spike}},\ }\href@noop {} {\  (\bibinfo {year} {2023})},\
  \Eprint {https://arxiv.org/abs/2312.14041} {arXiv:2312.14041 [gr-qc]}
  \BibitemShut {NoStop}%
\bibitem [{\citenamefont {Barausse}(2007)}]{Barausse:2007ph}%
  \BibitemOpen
  \bibfield  {author} {\bibinfo {author} {\bibfnamefont {E.}~\bibnamefont
  {Barausse}},\ }\bibfield  {title} {\bibinfo {title} {{Relativistic dynamical
  friction in a collisional fluid}},\ }\href
  {https://doi.org/10.1111/j.1365-2966.2007.12408.x} {\bibfield  {journal}
  {\bibinfo  {journal} {Mon. Not. Roy. Astron. Soc.}\ }\textbf {\bibinfo
  {volume} {382}},\ \bibinfo {pages} {826} (\bibinfo {year} {2007})},\ \Eprint
  {https://arxiv.org/abs/0709.0211} {arXiv:0709.0211 [astro-ph]} \BibitemShut
  {NoStop}%
\bibitem [{\citenamefont {Cardoso}\ \emph
  {et~al.}(2022{\natexlab{a}})\citenamefont {Cardoso}, \citenamefont
  {Destounis}, \citenamefont {Duque}, \citenamefont {Macedo},\ and\
  \citenamefont {Maselli}}]{Cardoso:2021wlq}%
  \BibitemOpen
  \bibfield  {author} {\bibinfo {author} {\bibfnamefont {V.}~\bibnamefont
  {Cardoso}}, \bibinfo {author} {\bibfnamefont {K.}~\bibnamefont {Destounis}},
  \bibinfo {author} {\bibfnamefont {F.}~\bibnamefont {Duque}}, \bibinfo
  {author} {\bibfnamefont {R.~P.}\ \bibnamefont {Macedo}},\ and\ \bibinfo
  {author} {\bibfnamefont {A.}~\bibnamefont {Maselli}},\ }\bibfield  {title}
  {\bibinfo {title} {{Black holes in galaxies: Environmental impact on
  gravitational-wave generation and propagation}},\ }\href
  {https://doi.org/10.1103/PhysRevD.105.L061501} {\bibfield  {journal}
  {\bibinfo  {journal} {Phys. Rev. D}\ }\textbf {\bibinfo {volume} {105}},\
  \bibinfo {pages} {L061501} (\bibinfo {year} {2022}{\natexlab{a}})},\ \Eprint
  {https://arxiv.org/abs/2109.00005} {arXiv:2109.00005 [gr-qc]} \BibitemShut
  {NoStop}%
\bibitem [{\citenamefont {Vicente}\ \emph {et~al.}(2025)\citenamefont
  {Vicente}, \citenamefont {Karydas},\ and\ \citenamefont
  {Bertone}}]{Vicente:2025gsg}%
  \BibitemOpen
  \bibfield  {author} {\bibinfo {author} {\bibfnamefont {R.}~\bibnamefont
  {Vicente}}, \bibinfo {author} {\bibfnamefont {T.~K.}\ \bibnamefont
  {Karydas}},\ and\ \bibinfo {author} {\bibfnamefont {G.}~\bibnamefont
  {Bertone}},\ }\bibfield  {title} {\bibinfo {title} {{A fully relativistic
  treatment of EMRIs in collisionless environments}},\ }\href@noop {} {\
  (\bibinfo {year} {2025})},\ \Eprint {https://arxiv.org/abs/2505.09715}
  {arXiv:2505.09715 [gr-qc]} \BibitemShut {NoStop}%
\bibitem [{\citenamefont {Khalvati}\ \emph {et~al.}(2024)\citenamefont
  {Khalvati}, \citenamefont {Santini}, \citenamefont {Duque}, \citenamefont
  {Speri}, \citenamefont {Gair}, \citenamefont {Yang},\ and\ \citenamefont
  {Brito}}]{Khalvati:2024tzz}%
  \BibitemOpen
  \bibfield  {author} {\bibinfo {author} {\bibfnamefont {H.}~\bibnamefont
  {Khalvati}}, \bibinfo {author} {\bibfnamefont {A.}~\bibnamefont {Santini}},
  \bibinfo {author} {\bibfnamefont {F.}~\bibnamefont {Duque}}, \bibinfo
  {author} {\bibfnamefont {L.}~\bibnamefont {Speri}}, \bibinfo {author}
  {\bibfnamefont {J.}~\bibnamefont {Gair}}, \bibinfo {author} {\bibfnamefont
  {H.}~\bibnamefont {Yang}},\ and\ \bibinfo {author} {\bibfnamefont
  {R.}~\bibnamefont {Brito}},\ }\bibfield  {title} {\bibinfo {title} {{Impact
  of relativistic waveforms in LISA's science objectives with
  extreme-mass-ratio inspirals}},\ }\href@noop {} {\  (\bibinfo {year}
  {2024})},\ \Eprint {https://arxiv.org/abs/2410.17310} {arXiv:2410.17310
  [gr-qc]} \BibitemShut {NoStop}%
\bibitem [{\citenamefont {Brito}\ and\ \citenamefont
  {Shah}(2023)}]{Brito:2023pyl}%
  \BibitemOpen
  \bibfield  {author} {\bibinfo {author} {\bibfnamefont {R.}~\bibnamefont
  {Brito}}\ and\ \bibinfo {author} {\bibfnamefont {S.}~\bibnamefont {Shah}},\
  }\bibfield  {title} {\bibinfo {title} {{Extreme mass-ratio inspirals into
  black holes surrounded by scalar clouds}},\ }\href@noop {} {\  (\bibinfo
  {year} {2023})},\ \Eprint {https://arxiv.org/abs/2307.16093}
  {arXiv:2307.16093 [gr-qc]} \BibitemShut {NoStop}%
\bibitem [{\citenamefont {Cardoso}\ \emph
  {et~al.}(2022{\natexlab{b}})\citenamefont {Cardoso}, \citenamefont
  {Destounis}, \citenamefont {Duque}, \citenamefont {Panosso~Macedo},\ and\
  \citenamefont {Maselli}}]{Cardoso:2022whc}%
  \BibitemOpen
  \bibfield  {author} {\bibinfo {author} {\bibfnamefont {V.}~\bibnamefont
  {Cardoso}}, \bibinfo {author} {\bibfnamefont {K.}~\bibnamefont {Destounis}},
  \bibinfo {author} {\bibfnamefont {F.}~\bibnamefont {Duque}}, \bibinfo
  {author} {\bibfnamefont {R.}~\bibnamefont {Panosso~Macedo}},\ and\ \bibinfo
  {author} {\bibfnamefont {A.}~\bibnamefont {Maselli}},\ }\bibfield  {title}
  {\bibinfo {title} {{Gravitational Waves from Extreme-Mass-Ratio Systems in
  Astrophysical Environments}},\ }\href
  {https://doi.org/10.1103/PhysRevLett.129.241103} {\bibfield  {journal}
  {\bibinfo  {journal} {Phys. Rev. Lett.}\ }\textbf {\bibinfo {volume} {129}},\
  \bibinfo {pages} {241103} (\bibinfo {year} {2022}{\natexlab{b}})},\ \Eprint
  {https://arxiv.org/abs/2210.01133} {arXiv:2210.01133 [gr-qc]} \BibitemShut
  {NoStop}%
\bibitem [{\citenamefont {Datta}(2022)}]{Datta:2021hvm}%
  \BibitemOpen
  \bibfield  {author} {\bibinfo {author} {\bibfnamefont {S.}~\bibnamefont
  {Datta}},\ }\bibfield  {title} {\bibinfo {title} {{Probing horizon scale
  quantum effects with Love}},\ }\href
  {https://doi.org/10.1088/1361-6382/ac9ae4} {\bibfield  {journal} {\bibinfo
  {journal} {Class. Quant. Grav.}\ }\textbf {\bibinfo {volume} {39}},\ \bibinfo
  {pages} {225016} (\bibinfo {year} {2022})},\ \Eprint
  {https://arxiv.org/abs/2107.07258} {arXiv:2107.07258 [gr-qc]} \BibitemShut
  {NoStop}%
\bibitem [{\citenamefont {Duque}\ \emph {et~al.}(2023)\citenamefont {Duque},
  \citenamefont {Macedo}, \citenamefont {Vicente},\ and\ \citenamefont
  {Cardoso}}]{Duque:2023cac}%
  \BibitemOpen
  \bibfield  {author} {\bibinfo {author} {\bibfnamefont {F.}~\bibnamefont
  {Duque}}, \bibinfo {author} {\bibfnamefont {C.~F.~B.}\ \bibnamefont
  {Macedo}}, \bibinfo {author} {\bibfnamefont {R.}~\bibnamefont {Vicente}},\
  and\ \bibinfo {author} {\bibfnamefont {V.}~\bibnamefont {Cardoso}},\
  }\bibfield  {title} {\bibinfo {title} {{Axion Weak Leaks: extreme mass-ratio
  inspirals in ultra-light dark matter}},\ }\href@noop {} {\  (\bibinfo {year}
  {2023})},\ \Eprint {https://arxiv.org/abs/2312.06767} {arXiv:2312.06767
  [gr-qc]} \BibitemShut {NoStop}%
\bibitem [{\citenamefont {Dyson}\ \emph {et~al.}(2025)\citenamefont {Dyson},
  \citenamefont {Spieksma}, \citenamefont {Brito}, \citenamefont {van~de
  Meent},\ and\ \citenamefont {Dolan}}]{dyson2025environmental}%
  \BibitemOpen
  \bibfield  {author} {\bibinfo {author} {\bibfnamefont {C.}~\bibnamefont
  {Dyson}}, \bibinfo {author} {\bibfnamefont {T.~F.}\ \bibnamefont {Spieksma}},
  \bibinfo {author} {\bibfnamefont {R.}~\bibnamefont {Brito}}, \bibinfo
  {author} {\bibfnamefont {M.}~\bibnamefont {van~de Meent}},\ and\ \bibinfo
  {author} {\bibfnamefont {S.}~\bibnamefont {Dolan}},\ }\bibfield  {title}
  {\bibinfo {title} {Environmental effects in extreme mass ratio inspirals:
  perturbations to the environment in kerr},\ }\href@noop {} {\bibfield
  {journal} {\bibinfo  {journal} {arXiv preprint arXiv:2501.09806}\ } (\bibinfo
  {year} {2025})}\BibitemShut {NoStop}%
\bibitem [{\citenamefont {Polcar}\ and\ \citenamefont
  {Witzany}(2025)}]{Polcar:2025yto}%
  \BibitemOpen
  \bibfield  {author} {\bibinfo {author} {\bibfnamefont {L.}~\bibnamefont
  {Polcar}}\ and\ \bibinfo {author} {\bibfnamefont {V.}~\bibnamefont
  {Witzany}},\ }\bibfield  {title} {\bibinfo {title} {{Towards relativistic
  inspirals into black holes surrounded by matter}},\ }\href@noop {} {\
  (\bibinfo {year} {2025})},\ \Eprint {https://arxiv.org/abs/2507.15720}
  {arXiv:2507.15720 [gr-qc]} \BibitemShut {NoStop}%
\bibitem [{\citenamefont {Xin}\ and\ \citenamefont {Most}(2025)}]{Xin:2025ymm}%
  \BibitemOpen
  \bibfield  {author} {\bibinfo {author} {\bibfnamefont {S.}~\bibnamefont
  {Xin}}\ and\ \bibinfo {author} {\bibfnamefont {E.~R.}\ \bibnamefont {Most}},\
  }\bibfield  {title} {\bibinfo {title} {{Relativistic scalar dark matter drag
  forces on a black hole binary}},\ }\href@noop {} {\  (\bibinfo {year}
  {2025})},\ \Eprint {https://arxiv.org/abs/2507.18934} {arXiv:2507.18934
  [gr-qc]} \BibitemShut {NoStop}%
\bibitem [{\citenamefont {Guo}\ \emph {et~al.}(2025)\citenamefont {Guo},
  \citenamefont {Zhong}, \citenamefont {Chen}, \citenamefont {Cardoso},
  \citenamefont {Ikeda},\ and\ \citenamefont {Zhou}}]{Guo:2025pea}%
  \BibitemOpen
  \bibfield  {author} {\bibinfo {author} {\bibfnamefont {Y.}~\bibnamefont
  {Guo}}, \bibinfo {author} {\bibfnamefont {Z.}~\bibnamefont {Zhong}}, \bibinfo
  {author} {\bibfnamefont {Y.}~\bibnamefont {Chen}}, \bibinfo {author}
  {\bibfnamefont {V.}~\bibnamefont {Cardoso}}, \bibinfo {author} {\bibfnamefont
  {T.}~\bibnamefont {Ikeda}},\ and\ \bibinfo {author} {\bibfnamefont
  {L.}~\bibnamefont {Zhou}},\ }\bibfield  {title} {\bibinfo {title}
  {{Ultralight Boson Ionization from Comparable-Mass Binary Black Holes}},\
  }\href@noop {} {\  (\bibinfo {year} {2025})},\ \Eprint
  {https://arxiv.org/abs/2509.09643} {arXiv:2509.09643 [gr-qc]} \BibitemShut
  {NoStop}%
\bibitem [{\citenamefont {Aurrekoetxea}\ \emph {et~al.}(2024)\citenamefont
  {Aurrekoetxea}, \citenamefont {Marsden}, \citenamefont {Clough},\ and\
  \citenamefont {Ferreira}}]{Aurrekoetxea:2024cqd}%
  \BibitemOpen
  \bibfield  {author} {\bibinfo {author} {\bibfnamefont {J.~C.}\ \bibnamefont
  {Aurrekoetxea}}, \bibinfo {author} {\bibfnamefont {J.}~\bibnamefont
  {Marsden}}, \bibinfo {author} {\bibfnamefont {K.}~\bibnamefont {Clough}},\
  and\ \bibinfo {author} {\bibfnamefont {P.~G.}\ \bibnamefont {Ferreira}},\
  }\bibfield  {title} {\bibinfo {title} {{Self-interacting scalar dark matter
  around binary black holes}},\ }\href
  {https://doi.org/10.1103/PhysRevD.110.083011} {\bibfield  {journal} {\bibinfo
   {journal} {Phys. Rev. D}\ }\textbf {\bibinfo {volume} {110}},\ \bibinfo
  {pages} {083011} (\bibinfo {year} {2024})},\ \Eprint
  {https://arxiv.org/abs/2409.01937} {arXiv:2409.01937 [gr-qc]} \BibitemShut
  {NoStop}%
\bibitem [{\citenamefont {Babak}\ \emph {et~al.}(2017)\citenamefont {Babak},
  \citenamefont {Gair}, \citenamefont {Sesana}, \citenamefont {Barausse},
  \citenamefont {Sopuerta}, \citenamefont {Berry}, \citenamefont {Berti},
  \citenamefont {Amaro-Seoane}, \citenamefont {Petiteau},\ and\ \citenamefont
  {Klein}}]{Babak:2017tow}%
  \BibitemOpen
  \bibfield  {author} {\bibinfo {author} {\bibfnamefont {S.}~\bibnamefont
  {Babak}}, \bibinfo {author} {\bibfnamefont {J.}~\bibnamefont {Gair}},
  \bibinfo {author} {\bibfnamefont {A.}~\bibnamefont {Sesana}}, \bibinfo
  {author} {\bibfnamefont {E.}~\bibnamefont {Barausse}}, \bibinfo {author}
  {\bibfnamefont {C.~F.}\ \bibnamefont {Sopuerta}}, \bibinfo {author}
  {\bibfnamefont {C.~P.~L.}\ \bibnamefont {Berry}}, \bibinfo {author}
  {\bibfnamefont {E.}~\bibnamefont {Berti}}, \bibinfo {author} {\bibfnamefont
  {P.}~\bibnamefont {Amaro-Seoane}}, \bibinfo {author} {\bibfnamefont
  {A.}~\bibnamefont {Petiteau}},\ and\ \bibinfo {author} {\bibfnamefont
  {A.}~\bibnamefont {Klein}},\ }\bibfield  {title} {\bibinfo {title} {{Science
  with the space-based interferometer LISA. V: Extreme mass-ratio inspirals}},\
  }\href {https://doi.org/10.1103/PhysRevD.95.103012} {\bibfield  {journal}
  {\bibinfo  {journal} {Phys. Rev.}\ }\textbf {\bibinfo {volume} {D95}},\
  \bibinfo {pages} {103012} (\bibinfo {year} {2017})},\ \Eprint
  {https://arxiv.org/abs/1703.09722} {arXiv:1703.09722 [gr-qc]} \BibitemShut
  {NoStop}%
\bibitem [{\citenamefont {Mancieri}\ \emph {et~al.}(2025)\citenamefont
  {Mancieri}, \citenamefont {Broggi}, \citenamefont {Vinciguerra},
  \citenamefont {Sesana},\ and\ \citenamefont {Bonetti}}]{Mancieri:2025cmx}%
  \BibitemOpen
  \bibfield  {author} {\bibinfo {author} {\bibfnamefont {D.}~\bibnamefont
  {Mancieri}}, \bibinfo {author} {\bibfnamefont {L.}~\bibnamefont {Broggi}},
  \bibinfo {author} {\bibfnamefont {M.}~\bibnamefont {Vinciguerra}}, \bibinfo
  {author} {\bibfnamefont {A.}~\bibnamefont {Sesana}},\ and\ \bibinfo {author}
  {\bibfnamefont {M.}~\bibnamefont {Bonetti}},\ }\bibfield  {title} {\bibinfo
  {title} {{Eccentricity distribution of extreme mass ratio inspirals}},\
  }\href@noop {} {\  (\bibinfo {year} {2025})},\ \Eprint
  {https://arxiv.org/abs/2509.02394} {arXiv:2509.02394 [astro-ph.HE]}
  \BibitemShut {NoStop}%
\bibitem [{\citenamefont {Colpi}\ \emph {et~al.}(2024)\citenamefont {Colpi}
  \emph {et~al.}}]{Colpi:2024xhw}%
  \BibitemOpen
  \bibfield  {author} {\bibinfo {author} {\bibfnamefont {M.}~\bibnamefont
  {Colpi}} \emph {et~al.},\ }\bibfield  {title} {\bibinfo {title} {{LISA
  Definition Study Report}},\ }\href@noop {} {\  (\bibinfo {year} {2024})},\
  \Eprint {https://arxiv.org/abs/2402.07571} {arXiv:2402.07571 [astro-ph.CO]}
  \BibitemShut {NoStop}%
\bibitem [{\citenamefont {Gong}\ \emph {et~al.}(2021)\citenamefont {Gong},
  \citenamefont {Luo},\ and\ \citenamefont {Wang}}]{Gong:2021gvw}%
  \BibitemOpen
  \bibfield  {author} {\bibinfo {author} {\bibfnamefont {Y.}~\bibnamefont
  {Gong}}, \bibinfo {author} {\bibfnamefont {J.}~\bibnamefont {Luo}},\ and\
  \bibinfo {author} {\bibfnamefont {B.}~\bibnamefont {Wang}},\ }\bibfield
  {title} {\bibinfo {title} {{Concepts and status of Chinese space
  gravitational wave detection projects}},\ }\href
  {https://doi.org/10.1038/s41550-021-01480-3} {\bibfield  {journal} {\bibinfo
  {journal} {Nature Astron.}\ }\textbf {\bibinfo {volume} {5}},\ \bibinfo
  {pages} {881} (\bibinfo {year} {2021})},\ \Eprint
  {https://arxiv.org/abs/2109.07442} {arXiv:2109.07442 [astro-ph.IM]}
  \BibitemShut {NoStop}%
\bibitem [{\citenamefont {Pound}\ and\ \citenamefont
  {Wardell}(2021{\natexlab{a}})}]{Pound:2021qin}%
  \BibitemOpen
  \bibfield  {author} {\bibinfo {author} {\bibfnamefont {A.}~\bibnamefont
  {Pound}}\ and\ \bibinfo {author} {\bibfnamefont {B.}~\bibnamefont
  {Wardell}},\ }\bibfield  {title} {\bibinfo {title} {{Black hole perturbation
  theory and gravitational self-force}}\ }\href
  {https://doi.org/10.1007/978-981-15-4702-7\_38-1}
  {10.1007/978-981-15-4702-7\_38-1} (\bibinfo {year} {2021}{\natexlab{a}}),\
  \Eprint {https://arxiv.org/abs/2101.04592} {arXiv:2101.04592 [gr-qc]}
  \BibitemShut {NoStop}%
\bibitem [{\citenamefont {Pound}\ and\ \citenamefont
  {Wardell}(2021{\natexlab{b}})}]{wardellpoundreview2021}%
  \BibitemOpen
  \bibfield  {author} {\bibinfo {author} {\bibfnamefont {A.}~\bibnamefont
  {Pound}}\ and\ \bibinfo {author} {\bibfnamefont {B.}~\bibnamefont
  {Wardell}},\ }\bibfield  {title} {\bibinfo {title} {Black hole perturbation
  theory and gravitational self-force},\ }\href@noop {} {\bibfield  {journal}
  {\bibinfo  {journal} {arXiv preprint arXiv:2101.04592}\ } (\bibinfo {year}
  {2021}{\natexlab{b}})}\BibitemShut {NoStop}%
\bibitem [{\citenamefont {Duque}\ \emph {et~al.}(2025)\citenamefont {Duque},
  \citenamefont {Kejriwal}, \citenamefont {Sberna}, \citenamefont {Speri},\
  and\ \citenamefont {Gair}}]{Duque:2024mfw}%
  \BibitemOpen
  \bibfield  {author} {\bibinfo {author} {\bibfnamefont {F.}~\bibnamefont
  {Duque}}, \bibinfo {author} {\bibfnamefont {S.}~\bibnamefont {Kejriwal}},
  \bibinfo {author} {\bibfnamefont {L.}~\bibnamefont {Sberna}}, \bibinfo
  {author} {\bibfnamefont {L.}~\bibnamefont {Speri}},\ and\ \bibinfo {author}
  {\bibfnamefont {J.}~\bibnamefont {Gair}},\ }\bibfield  {title} {\bibinfo
  {title} {{Constraining accretion physics with gravitational waves from
  eccentric extreme-mass-ratio inspirals}},\ }\href
  {https://doi.org/10.1103/PhysRevD.111.084006} {\bibfield  {journal} {\bibinfo
   {journal} {Phys. Rev. D}\ }\textbf {\bibinfo {volume} {111}},\ \bibinfo
  {pages} {084006} (\bibinfo {year} {2025})},\ \Eprint
  {https://arxiv.org/abs/2411.03436} {arXiv:2411.03436 [gr-qc]} \BibitemShut
  {NoStop}%
\bibitem [{\citenamefont {Sun}\ \emph {et~al.}(2025)\citenamefont {Sun},
  \citenamefont {Li}, \citenamefont {Pan},\ and\ \citenamefont
  {Yang}}]{Sun:2025lbr}%
  \BibitemOpen
  \bibfield  {author} {\bibinfo {author} {\bibfnamefont {H.}~\bibnamefont
  {Sun}}, \bibinfo {author} {\bibfnamefont {Y.-P.}\ \bibnamefont {Li}},
  \bibinfo {author} {\bibfnamefont {Z.}~\bibnamefont {Pan}},\ and\ \bibinfo
  {author} {\bibfnamefont {H.}~\bibnamefont {Yang}},\ }\bibfield  {title}
  {\bibinfo {title} {{Probing Formation Channels of Extreme Mass-Ratio
  Inspirals}},\ }\href@noop {} {\  (\bibinfo {year} {2025})},\ \Eprint
  {https://arxiv.org/abs/2509.00469} {arXiv:2509.00469 [gr-qc]} \BibitemShut
  {NoStop}%
\bibitem [{\citenamefont {Franchini}\ \emph {et~al.}(2023)\citenamefont
  {Franchini}, \citenamefont {Bonetti}, \citenamefont {Lupi}, \citenamefont
  {Miniutti}, \citenamefont {Bortolas}, \citenamefont {Giustini}, \citenamefont
  {Dotti}, \citenamefont {Sesana}, \citenamefont {Arcodia},\ and\ \citenamefont
  {Ryu}}]{Franchini:2023bou}%
  \BibitemOpen
  \bibfield  {author} {\bibinfo {author} {\bibfnamefont {A.}~\bibnamefont
  {Franchini}}, \bibinfo {author} {\bibfnamefont {M.}~\bibnamefont {Bonetti}},
  \bibinfo {author} {\bibfnamefont {A.}~\bibnamefont {Lupi}}, \bibinfo {author}
  {\bibfnamefont {G.}~\bibnamefont {Miniutti}}, \bibinfo {author}
  {\bibfnamefont {E.}~\bibnamefont {Bortolas}}, \bibinfo {author}
  {\bibfnamefont {M.}~\bibnamefont {Giustini}}, \bibinfo {author}
  {\bibfnamefont {M.}~\bibnamefont {Dotti}}, \bibinfo {author} {\bibfnamefont
  {A.}~\bibnamefont {Sesana}}, \bibinfo {author} {\bibfnamefont
  {R.}~\bibnamefont {Arcodia}},\ and\ \bibinfo {author} {\bibfnamefont
  {T.}~\bibnamefont {Ryu}},\ }\bibfield  {title} {\bibinfo {title}
  {{Quasi-periodic eruptions from impacts between the secondary and a rigidly
  precessing accretion disc in an extreme mass-ratio inspiral system}},\ }\href
  {https://doi.org/10.1051/0004-6361/202346565} {\bibfield  {journal} {\bibinfo
   {journal} {Astron. Astrophys.}\ }\textbf {\bibinfo {volume} {675}},\
  \bibinfo {pages} {A100} (\bibinfo {year} {2023})},\ \Eprint
  {https://arxiv.org/abs/2304.00775} {arXiv:2304.00775 [astro-ph.HE]}
  \BibitemShut {NoStop}%
\bibitem [{\citenamefont {Linial}\ and\ \citenamefont
  {Metzger}(2023)}]{Linial:2023nqs}%
  \BibitemOpen
  \bibfield  {author} {\bibinfo {author} {\bibfnamefont {I.}~\bibnamefont
  {Linial}}\ and\ \bibinfo {author} {\bibfnamefont {B.~D.}\ \bibnamefont
  {Metzger}},\ }\bibfield  {title} {\bibinfo {title} {{EMRI + TDE = QPE:
  Periodic X-Ray Flares from Star{\textendash}Disk Collisions in Galactic
  Nuclei}},\ }\href {https://doi.org/10.3847/1538-4357/acf65b} {\bibfield
  {journal} {\bibinfo  {journal} {Astrophys. J.}\ }\textbf {\bibinfo {volume}
  {957}},\ \bibinfo {pages} {34} (\bibinfo {year} {2023})},\ \Eprint
  {https://arxiv.org/abs/2303.16231} {arXiv:2303.16231 [astro-ph.HE]}
  \BibitemShut {NoStop}%
\bibitem [{\citenamefont {Chakraborty}\ \emph {et~al.}(2025)\citenamefont
  {Chakraborty} \emph {et~al.}}]{Chakraborty:2025xch}%
  \BibitemOpen
  \bibfield  {author} {\bibinfo {author} {\bibfnamefont {J.}~\bibnamefont
  {Chakraborty}} \emph {et~al.},\ }\bibfield  {title} {\bibinfo {title}
  {{Prospects for EMRI/MBH parameter estimation using Quasi-Periodic Eruption
  timings: short-timescale analysis}},\ }\href@noop {} {\  (\bibinfo {year}
  {2025})},\ \Eprint {https://arxiv.org/abs/2508.20162} {arXiv:2508.20162
  [astro-ph.HE]} \BibitemShut {NoStop}%
\bibitem [{\citenamefont {{Goldreich}}\ and\ \citenamefont
  {{Tremaine}}(1980)}]{GoldreichTremaine1980}%
  \BibitemOpen
  \bibfield  {author} {\bibinfo {author} {\bibfnamefont {P.}~\bibnamefont
  {{Goldreich}}}\ and\ \bibinfo {author} {\bibfnamefont {S.}~\bibnamefont
  {{Tremaine}}},\ }\bibfield  {title} {\bibinfo {title} {{Disk-satellite
  interactions.}},\ }\href {https://doi.org/10.1086/158356} {\bibfield
  {journal} {\bibinfo  {journal} {The Astrophysics Journal}\ }\textbf {\bibinfo
  {volume} {241}},\ \bibinfo {pages} {425} (\bibinfo {year}
  {1980})}\BibitemShut {NoStop}%
\bibitem [{\citenamefont {{Artymowicz}}(1993)}]{Artymowicz1993}%
  \BibitemOpen
  \bibfield  {author} {\bibinfo {author} {\bibfnamefont {P.}~\bibnamefont
  {{Artymowicz}}},\ }\bibfield  {title} {\bibinfo {title} {{On the Wave
  Excitation and a Generalized Torque Formula for Lindblad Resonances Excited
  by External Potential}},\ }\href {https://doi.org/10.1086/173469} {\bibfield
  {journal} {\bibinfo  {journal} {The Astrophysics Journal}\ }\textbf {\bibinfo
  {volume} {419}},\ \bibinfo {pages} {155} (\bibinfo {year}
  {1993})}\BibitemShut {NoStop}%
\bibitem [{\citenamefont {Tanaka}\ \emph {et~al.}(2002)\citenamefont {Tanaka},
  \citenamefont {Takeuchi},\ and\ \citenamefont {Ward}}]{Tanaka_2002}%
  \BibitemOpen
  \bibfield  {author} {\bibinfo {author} {\bibfnamefont {H.}~\bibnamefont
  {Tanaka}}, \bibinfo {author} {\bibfnamefont {T.}~\bibnamefont {Takeuchi}},\
  and\ \bibinfo {author} {\bibfnamefont {W.~R.}\ \bibnamefont {Ward}},\
  }\bibfield  {title} {\bibinfo {title} {Three-dimensional interaction between
  a planet and an isothermal gaseous disk. i. corotation and lindblad torques
  and planet migration},\ }\href {https://doi.org/10.1086/324713} {\bibfield
  {journal} {\bibinfo  {journal} {The Astrophysical Journal}\ }\textbf
  {\bibinfo {volume} {565}},\ \bibinfo {pages} {1257} (\bibinfo {year}
  {2002})}\BibitemShut {NoStop}%
\bibitem [{\citenamefont {Tanaka}\ and\ \citenamefont
  {Ward}(2004)}]{Tanaka_2004}%
  \BibitemOpen
  \bibfield  {author} {\bibinfo {author} {\bibfnamefont {H.}~\bibnamefont
  {Tanaka}}\ and\ \bibinfo {author} {\bibfnamefont {W.~R.}\ \bibnamefont
  {Ward}},\ }\bibfield  {title} {\bibinfo {title} {Three-dimensional
  interaction between a planet and an isothermal gaseous disk. ii. eccentricity
  waves and bending waves},\ }\href {https://doi.org/10.1086/380992} {\bibfield
   {journal} {\bibinfo  {journal} {The Astrophysical Journal}\ }\textbf
  {\bibinfo {volume} {602}},\ \bibinfo {pages} {388} (\bibinfo {year}
  {2004})}\BibitemShut {NoStop}%
\bibitem [{\citenamefont {{Tanaka}}\ and\ \citenamefont
  {{Okada}}(2024)}]{Tanaka3}%
  \BibitemOpen
  \bibfield  {author} {\bibinfo {author} {\bibfnamefont {H.}~\bibnamefont
  {{Tanaka}}}\ and\ \bibinfo {author} {\bibfnamefont {K.}~\bibnamefont
  {{Okada}}},\ }\bibfield  {title} {\bibinfo {title} {{Three-dimensional
  Interaction between a Planet and an Isothermal Gaseous Disk. III. Locally
  Isothermal Cases}},\ }\href {https://doi.org/10.3847/1538-4357/ad410d}
  {\bibfield  {journal} {\bibinfo  {journal} {The Astrophysics Journal}\
  }\textbf {\bibinfo {volume} {968}},\ \bibinfo {eid} {28} (\bibinfo {year}
  {2024})},\ \Eprint {https://arxiv.org/abs/2404.12521} {arXiv:2404.12521
  [astro-ph.EP]} \BibitemShut {NoStop}%
\bibitem [{\citenamefont {{Fairbairn}}\ and\ \citenamefont
  {{Rafikov}}(2025)}]{rafikov_eccentricity}%
  \BibitemOpen
  \bibfield  {author} {\bibinfo {author} {\bibfnamefont {C.~W.}\ \bibnamefont
  {{Fairbairn}}}\ and\ \bibinfo {author} {\bibfnamefont {R.~R.}\ \bibnamefont
  {{Rafikov}}},\ }\bibfield  {title} {\bibinfo {title} {{Eccentric planet-disc
  interactions: orbital migration and eccentricity evolution}},\ }\href
  {https://doi.org/10.1093/mnras/staf117} {\bibfield  {journal} {\bibinfo
  {journal} {MNRAS}\ }\textbf {\bibinfo {volume} {537}},\ \bibinfo {pages}
  {1779} (\bibinfo {year} {2025})},\ \Eprint {https://arxiv.org/abs/2407.20398}
  {arXiv:2407.20398 [astro-ph.EP]} \BibitemShut {NoStop}%
\bibitem [{\citenamefont {Yunes}\ \emph {et~al.}(2011)\citenamefont {Yunes},
  \citenamefont {Kocsis}, \citenamefont {Loeb},\ and\ \citenamefont
  {Haiman}}]{Yunes:2011ws}%
  \BibitemOpen
  \bibfield  {author} {\bibinfo {author} {\bibfnamefont {N.}~\bibnamefont
  {Yunes}}, \bibinfo {author} {\bibfnamefont {B.}~\bibnamefont {Kocsis}},
  \bibinfo {author} {\bibfnamefont {A.}~\bibnamefont {Loeb}},\ and\ \bibinfo
  {author} {\bibfnamefont {Z.}~\bibnamefont {Haiman}},\ }\bibfield  {title}
  {\bibinfo {title} {{Imprint of Accretion Disk-Induced Migration on
  Gravitational Waves from Extreme Mass Ratio Inspirals}},\ }\href
  {https://doi.org/10.1103/PhysRevLett.107.171103} {\bibfield  {journal}
  {\bibinfo  {journal} {Phys. Rev. Lett.}\ }\textbf {\bibinfo {volume} {107}},\
  \bibinfo {pages} {171103} (\bibinfo {year} {2011})},\ \Eprint
  {https://arxiv.org/abs/1103.4609} {arXiv:1103.4609 [astro-ph.CO]}
  \BibitemShut {NoStop}%
\bibitem [{\citenamefont {Kejriwal}\ \emph {et~al.}(2023)\citenamefont
  {Kejriwal}, \citenamefont {Speri},\ and\ \citenamefont
  {Chua}}]{Kejriwal:2023djc}%
  \BibitemOpen
  \bibfield  {author} {\bibinfo {author} {\bibfnamefont {S.}~\bibnamefont
  {Kejriwal}}, \bibinfo {author} {\bibfnamefont {L.}~\bibnamefont {Speri}},\
  and\ \bibinfo {author} {\bibfnamefont {A.~J.~K.}\ \bibnamefont {Chua}},\
  }\bibfield  {title} {\bibinfo {title} {{Impact of Correlations on the
  Modeling and Inference of Beyond Vacuum-GR Effects in Extreme-Mass-Ratio
  Inspirals}},\ }\href@noop {} {\  (\bibinfo {year} {2023})},\ \Eprint
  {https://arxiv.org/abs/2312.13028} {arXiv:2312.13028 [gr-qc]} \BibitemShut
  {NoStop}%
\bibitem [{\citenamefont {Hirata}(2011{\natexlab{a}})}]{Hirata:2010vn}%
  \BibitemOpen
  \bibfield  {author} {\bibinfo {author} {\bibfnamefont {C.~M.}\ \bibnamefont
  {Hirata}},\ }\bibfield  {title} {\bibinfo {title} {{Lindblad resonance
  torques in relativistic discs: I. Basic equations}},\ }\href
  {https://doi.org/10.1111/j.1365-2966.2011.18617.x} {\bibfield  {journal}
  {\bibinfo  {journal} {Mon. Not. Roy. Astron. Soc.}\ }\textbf {\bibinfo
  {volume} {414}},\ \bibinfo {pages} {3198} (\bibinfo {year}
  {2011}{\natexlab{a}})},\ \Eprint {https://arxiv.org/abs/1010.0758}
  {arXiv:1010.0758 [astro-ph.HE]} \BibitemShut {NoStop}%
\bibitem [{\citenamefont {Hirata}(2011{\natexlab{b}})}]{Hirata:2010vp}%
  \BibitemOpen
  \bibfield  {author} {\bibinfo {author} {\bibfnamefont {C.~M.}\ \bibnamefont
  {Hirata}},\ }\bibfield  {title} {\bibinfo {title} {{Lindblad resonance
  torques in relativistic discs: II. Computation of resonance strengths}},\
  }\href {https://doi.org/10.1111/j.1365-2966.2011.18619.x} {\bibfield
  {journal} {\bibinfo  {journal} {Mon. Not. Roy. Astron. Soc.}\ }\textbf
  {\bibinfo {volume} {414}},\ \bibinfo {pages} {3212} (\bibinfo {year}
  {2011}{\natexlab{b}})},\ \Eprint {https://arxiv.org/abs/1010.0759}
  {arXiv:1010.0759 [astro-ph.HE]} \BibitemShut {NoStop}%
\bibitem [{\citenamefont {Hegade K.~R.}\ \emph
  {et~al.}(2025{\natexlab{a}})\citenamefont {Hegade K.~R.}, \citenamefont
  {Gammie},\ and\ \citenamefont {Yunes}}]{HegadeKR:2025dur}%
  \BibitemOpen
  \bibfield  {author} {\bibinfo {author} {\bibfnamefont {A.}~\bibnamefont
  {Hegade K.~R.}}, \bibinfo {author} {\bibfnamefont {C.~F.}\ \bibnamefont
  {Gammie}},\ and\ \bibinfo {author} {\bibfnamefont {N.}~\bibnamefont
  {Yunes}},\ }\bibfield  {title} {\bibinfo {title} {{A relativistic treatment
  of accretion disk torques on extreme mass-ratio inspirals around non-spinning
  black holes}},\ }\href@noop {} {\  (\bibinfo {year} {2025}{\natexlab{a}})},\
  \Eprint {https://arxiv.org/abs/2509.20457} {arXiv:2509.20457 [gr-qc]}
  \BibitemShut {NoStop}%
\bibitem [{\citenamefont {Hegade K.~R.}\ \emph
  {et~al.}(2025{\natexlab{b}})\citenamefont {Hegade K.~R.}, \citenamefont
  {Gammie},\ and\ \citenamefont {Yunes}}]{HegadeKR:2025rpr}%
  \BibitemOpen
  \bibfield  {author} {\bibinfo {author} {\bibfnamefont {A.}~\bibnamefont
  {Hegade K.~R.}}, \bibinfo {author} {\bibfnamefont {C.~F.}\ \bibnamefont
  {Gammie}},\ and\ \bibinfo {author} {\bibfnamefont {N.}~\bibnamefont
  {Yunes}},\ }\bibfield  {title} {\bibinfo {title} {{A relativistic treatment
  of accretion disk torques on extreme mass ratio inspirals around spinning
  black holes}},\ }\href@noop {} {\  (\bibinfo {year} {2025}{\natexlab{b}})},\
  \Eprint {https://arxiv.org/abs/2510.03564} {arXiv:2510.03564 [gr-qc]}
  \BibitemShut {NoStop}%
\bibitem [{\citenamefont {Bardeen}\ and\ \citenamefont
  {Petterson}(1975)}]{Bardeen:1975zz}%
  \BibitemOpen
  \bibfield  {author} {\bibinfo {author} {\bibfnamefont {J.~M.}\ \bibnamefont
  {Bardeen}}\ and\ \bibinfo {author} {\bibfnamefont {J.~A.}\ \bibnamefont
  {Petterson}},\ }\bibfield  {title} {\bibinfo {title} {{The Lense-Thirring
  Effect and Accretion Disks around Kerr Black Holes}},\ }\href
  {https://doi.org/10.1086/181711} {\bibfield  {journal} {\bibinfo  {journal}
  {Astrophys. J. Lett.}\ }\textbf {\bibinfo {volume} {195}},\ \bibinfo {pages}
  {L65} (\bibinfo {year} {1975})}\BibitemShut {NoStop}%
\bibitem [{\citenamefont {Natarajan}\ and\ \citenamefont
  {Pringle}(1998)}]{Natarajan:1998xt}%
  \BibitemOpen
  \bibfield  {author} {\bibinfo {author} {\bibfnamefont {P.}~\bibnamefont
  {Natarajan}}\ and\ \bibinfo {author} {\bibfnamefont {J.~E.}\ \bibnamefont
  {Pringle}},\ }\bibfield  {title} {\bibinfo {title} {{The Alignment of disk
  and black hole spins in active galactic nuclei}},\ }\href
  {https://doi.org/10.1086/311658} {\bibfield  {journal} {\bibinfo  {journal}
  {Astrophys. J. Lett.}\ }\textbf {\bibinfo {volume} {506}},\ \bibinfo {pages}
  {L97} (\bibinfo {year} {1998})},\ \Eprint
  {https://arxiv.org/abs/astro-ph/9808187} {arXiv:astro-ph/9808187}
  \BibitemShut {NoStop}%
\bibitem [{\citenamefont {Chatterjee}\ \emph {et~al.}(2025)\citenamefont
  {Chatterjee}, \citenamefont {Kaaz}, \citenamefont {Liska}, \citenamefont
  {Tchekhovskoy},\ and\ \citenamefont {Markoff}}]{Chatterjee:2023ber}%
  \BibitemOpen
  \bibfield  {author} {\bibinfo {author} {\bibfnamefont {K.}~\bibnamefont
  {Chatterjee}}, \bibinfo {author} {\bibfnamefont {N.}~\bibnamefont {Kaaz}},
  \bibinfo {author} {\bibfnamefont {M.}~\bibnamefont {Liska}}, \bibinfo
  {author} {\bibfnamefont {A.}~\bibnamefont {Tchekhovskoy}},\ and\ \bibinfo
  {author} {\bibfnamefont {S.}~\bibnamefont {Markoff}},\ }\bibfield  {title}
  {\bibinfo {title} {{Misaligned magnetized accretion flows onto spinning black
  holes: Magneto-spin alignment, outflow power, and intermittent jets}},\
  }\href {https://doi.org/10.1103/hgj9-v4fk} {\bibfield  {journal} {\bibinfo
  {journal} {Phys. Rev. D}\ }\textbf {\bibinfo {volume} {112}},\ \bibinfo
  {pages} {063013} (\bibinfo {year} {2025})},\ \Eprint
  {https://arxiv.org/abs/2311.00432} {arXiv:2311.00432 [astro-ph.HE]}
  \BibitemShut {NoStop}%
\bibitem [{\citenamefont {Lobban}\ and\ \citenamefont
  {King}(2022)}]{Lobban:2022aon}%
  \BibitemOpen
  \bibfield  {author} {\bibinfo {author} {\bibfnamefont {A.}~\bibnamefont
  {Lobban}}\ and\ \bibinfo {author} {\bibfnamefont {A.}~\bibnamefont {King}},\
  }\bibfield  {title} {\bibinfo {title} {{AGN light echoes and the accretion
  disc self-gravity limit}},\ }\href {https://doi.org/10.1093/mnras/stac155}
  {\bibfield  {journal} {\bibinfo  {journal} {Mon. Not. Roy. Astron. Soc.}\
  }\textbf {\bibinfo {volume} {511}},\ \bibinfo {pages} {1992} (\bibinfo {year}
  {2022})},\ \Eprint {https://arxiv.org/abs/2201.08375} {arXiv:2201.08375
  [astro-ph.GA]} \BibitemShut {NoStop}%
\bibitem [{\citenamefont {Spieksma}\ and\ \citenamefont
  {Cannizzaro}(2025)}]{Spieksma:2025wex}%
  \BibitemOpen
  \bibfield  {author} {\bibinfo {author} {\bibfnamefont {T.~F.~M.}\
  \bibnamefont {Spieksma}}\ and\ \bibinfo {author} {\bibfnamefont
  {E.}~\bibnamefont {Cannizzaro}},\ }\bibfield  {title} {\bibinfo {title} {{In
  the grip of the disk: dragging the companion through an AGN}},\ }\href@noop
  {} {\  (\bibinfo {year} {2025})},\ \Eprint {https://arxiv.org/abs/2504.08033}
  {arXiv:2504.08033 [astro-ph.GA]} \BibitemShut {NoStop}%
\bibitem [{\citenamefont {{Wang}}\ \emph {et~al.}(2024)\citenamefont {{Wang}},
  \citenamefont {{Zhu}},\ and\ \citenamefont {{Lin}}}]{2024MNRAS.528.4958W}%
  \BibitemOpen
  \bibfield  {author} {\bibinfo {author} {\bibfnamefont {Y.}~\bibnamefont
  {{Wang}}}, \bibinfo {author} {\bibfnamefont {Z.}~\bibnamefont {{Zhu}}},\ and\
  \bibinfo {author} {\bibfnamefont {D.~N.~C.}\ \bibnamefont {{Lin}}},\
  }\bibfield  {title} {\bibinfo {title} {{Stellar/BH population in AGN discs:
  direct binary formation from capture objects in nuclei clusters}},\ }\href
  {https://doi.org/10.1093/mnras/stae321} {\bibfield  {journal} {\bibinfo
  {journal} {MNRAS}\ }\textbf {\bibinfo {volume} {528}},\ \bibinfo {pages}
  {4958} (\bibinfo {year} {2024})},\ \Eprint {https://arxiv.org/abs/2308.09129}
  {arXiv:2308.09129 [astro-ph.GA]} \BibitemShut {NoStop}%
\bibitem [{\citenamefont {Teukolsky}\ and\ \citenamefont
  {Press}(1974)}]{teukolsky1974perturbations}%
  \BibitemOpen
  \bibfield  {author} {\bibinfo {author} {\bibfnamefont {S.~A.}\ \bibnamefont
  {Teukolsky}}\ and\ \bibinfo {author} {\bibfnamefont {W.}~\bibnamefont
  {Press}},\ }\bibfield  {title} {\bibinfo {title} {Perturbations of a rotating
  black hole. iii-interaction of the hole with gravitational and
  electromagnetic radiation},\ }\href@noop {} {\bibfield  {journal} {\bibinfo
  {journal} {The Astrophysical Journal}\ }\textbf {\bibinfo {volume} {193}},\
  \bibinfo {pages} {443} (\bibinfo {year} {1974})}\BibitemShut {NoStop}%
\bibitem [{\citenamefont {Kerr}(1963)}]{kerr1963gravitational}%
  \BibitemOpen
  \bibfield  {author} {\bibinfo {author} {\bibfnamefont {R.~P.}\ \bibnamefont
  {Kerr}},\ }\bibfield  {title} {\bibinfo {title} {Gravitational field of a
  spinning mass as an example of algebraically special metrics},\ }\href@noop
  {} {\bibfield  {journal} {\bibinfo  {journal} {Physical review letters}\
  }\textbf {\bibinfo {volume} {11}},\ \bibinfo {pages} {237} (\bibinfo {year}
  {1963})}\BibitemShut {NoStop}%
\bibitem [{\citenamefont {Teukolsky}(1972)}]{teuk1972}%
  \BibitemOpen
  \bibfield  {author} {\bibinfo {author} {\bibfnamefont {S.~A.}\ \bibnamefont
  {Teukolsky}},\ }\bibfield  {title} {\bibinfo {title} {Rotating black holes:
  Separable wave equations for gravitational and electromagnetic
  perturbations},\ }\href@noop {} {\bibfield  {journal} {\bibinfo  {journal}
  {Physical Review Letters}\ }\textbf {\bibinfo {volume} {29}},\ \bibinfo
  {pages} {1114} (\bibinfo {year} {1972})}\BibitemShut {NoStop}%
\bibitem [{\citenamefont {Teukolsky}(1973)}]{teuk1973}%
  \BibitemOpen
  \bibfield  {author} {\bibinfo {author} {\bibfnamefont {S.~A.}\ \bibnamefont
  {Teukolsky}},\ }\bibfield  {title} {\bibinfo {title} {Perturbations of a
  rotating black hole. 1. fundamental equations for gravitational
  electromagnetic and neutrino field perturbations},\ }\href@noop {} {\bibfield
   {journal} {\bibinfo  {journal} {Astrophys. J.}\ }\textbf {\bibinfo {volume}
  {185}},\ \bibinfo {pages} {635} (\bibinfo {year} {1973})}\BibitemShut
  {NoStop}%
\bibitem [{\citenamefont {Chandrasekhar}(1992)}]{chandrabook}%
  \BibitemOpen
  \bibfield  {author} {\bibinfo {author} {\bibfnamefont {S.}~\bibnamefont
  {Chandrasekhar}},\ }\href@noop {} {\emph {\bibinfo {title} {The Mathematical
  Theory of Black Holes}}}\ (\bibinfo  {publisher} {Clarendon Press Oxford},\
  \bibinfo {year} {1992})\ \bibinfo {note} {reprint: 2009}\BibitemShut
  {NoStop}%
\bibitem [{\citenamefont {Hughes}\ \emph {et~al.}(2021)\citenamefont {Hughes},
  \citenamefont {Warburton}, \citenamefont {Khanna}, \citenamefont {Chua},\
  and\ \citenamefont {Katz}}]{Hughes:2021exa}%
  \BibitemOpen
  \bibfield  {author} {\bibinfo {author} {\bibfnamefont {S.~A.}\ \bibnamefont
  {Hughes}}, \bibinfo {author} {\bibfnamefont {N.}~\bibnamefont {Warburton}},
  \bibinfo {author} {\bibfnamefont {G.}~\bibnamefont {Khanna}}, \bibinfo
  {author} {\bibfnamefont {A.~J.~K.}\ \bibnamefont {Chua}},\ and\ \bibinfo
  {author} {\bibfnamefont {M.~L.}\ \bibnamefont {Katz}},\ }\bibfield  {title}
  {\bibinfo {title} {{Adiabatic waveforms for extreme mass-ratio inspirals via
  multivoice decomposition in time and frequency}},\ }\href
  {https://doi.org/10.1103/PhysRevD.103.104014} {\bibfield  {journal} {\bibinfo
   {journal} {Phys. Rev. D}\ }\textbf {\bibinfo {volume} {103}},\ \bibinfo
  {pages} {104014} (\bibinfo {year} {2021})},\ \Eprint
  {https://arxiv.org/abs/2102.02713} {arXiv:2102.02713 [gr-qc]} \BibitemShut
  {NoStop}%
\bibitem [{\citenamefont {Nasipak}(2025)}]{Nasipak:2025tby}%
  \BibitemOpen
  \bibfield  {author} {\bibinfo {author} {\bibfnamefont {Z.}~\bibnamefont
  {Nasipak}},\ }\bibfield  {title} {\bibinfo {title} {{Metric reconstruction
  and the Hamiltonian for eccentric, precessing binaries in the
  small-mass-ratio limit}},\ }\href@noop {} {\  (\bibinfo {year} {2025})},\
  \Eprint {https://arxiv.org/abs/2507.07746} {arXiv:2507.07746 [gr-qc]}
  \BibitemShut {NoStop}%
\bibitem [{\citenamefont {{Ward}}(1997)}]{Ward1997}%
  \BibitemOpen
  \bibfield  {author} {\bibinfo {author} {\bibfnamefont {W.~R.}\ \bibnamefont
  {{Ward}}},\ }\bibfield  {title} {\bibinfo {title} {{Protoplanet Migration by
  Nebula Tides}},\ }\href {https://doi.org/10.1006/icar.1996.5647} {\bibfield
  {journal} {\bibinfo  {journal} {Icarus}\ }\textbf {\bibinfo {volume} {126}},\
  \bibinfo {pages} {261} (\bibinfo {year} {1997})}\BibitemShut {NoStop}%
\bibitem [{\citenamefont {Gangardt}\ \emph {et~al.}(2024)\citenamefont
  {Gangardt}, \citenamefont {Trani}, \citenamefont {Bonnerot},\ and\
  \citenamefont {Gerosa}}]{Gangardt:2024bic}%
  \BibitemOpen
  \bibfield  {author} {\bibinfo {author} {\bibfnamefont {D.}~\bibnamefont
  {Gangardt}}, \bibinfo {author} {\bibfnamefont {A.~A.}\ \bibnamefont {Trani}},
  \bibinfo {author} {\bibfnamefont {C.}~\bibnamefont {Bonnerot}},\ and\
  \bibinfo {author} {\bibfnamefont {D.}~\bibnamefont {Gerosa}},\ }\bibfield
  {title} {\bibinfo {title} {{pAGN: the one-stop solution for AGN disc
  modeling}},\ }\href@noop {} {\  (\bibinfo {year} {2024})},\ \Eprint
  {https://arxiv.org/abs/2403.00060} {arXiv:2403.00060 [astro-ph.HE]}
  \BibitemShut {NoStop}%
\bibitem [{\citenamefont {{Goldreich}}\ and\ \citenamefont
  {{Tremaine}}(1979)}]{GoldreichTremaine1979A}%
  \BibitemOpen
  \bibfield  {author} {\bibinfo {author} {\bibfnamefont {P.}~\bibnamefont
  {{Goldreich}}}\ and\ \bibinfo {author} {\bibfnamefont {S.}~\bibnamefont
  {{Tremaine}}},\ }\bibfield  {title} {\bibinfo {title} {{The excitation of
  density waves at the Lindblad and corotation resonances by an external
  potential.}},\ }\href {https://doi.org/10.1086/157448} {\bibfield  {journal}
  {\bibinfo  {journal} {The Astrophysics Journal}\ }\textbf {\bibinfo {volume}
  {233}},\ \bibinfo {pages} {857} (\bibinfo {year} {1979})}\BibitemShut
  {NoStop}%
\bibitem [{\citenamefont {Novikov}\ and\ \citenamefont
  {Thorne}(1973)}]{Novikov:1973kta}%
  \BibitemOpen
  \bibfield  {author} {\bibinfo {author} {\bibfnamefont {I.~D.}\ \bibnamefont
  {Novikov}}\ and\ \bibinfo {author} {\bibfnamefont {K.~S.}\ \bibnamefont
  {Thorne}},\ }\bibfield  {title} {\bibinfo {title} {{Astrophysics and black
  holes}},\ }in\ \href@noop {} {\emph {\bibinfo {booktitle} {{Astrophysics of
  black holes}}}}\ (\bibinfo {year} {1973})\ pp.\ \bibinfo {pages}
  {343--550}\BibitemShut {NoStop}%
\bibitem [{\citenamefont {Shakura}\ and\ \citenamefont
  {Sunyaev}(1973)}]{Shakura:1972te}%
  \BibitemOpen
  \bibfield  {author} {\bibinfo {author} {\bibfnamefont {N.~I.}\ \bibnamefont
  {Shakura}}\ and\ \bibinfo {author} {\bibfnamefont {R.~A.}\ \bibnamefont
  {Sunyaev}},\ }\bibfield  {title} {\bibinfo {title} {{Black holes in binary
  systems. Observational appearance}},\ }\href@noop {} {\bibfield  {journal}
  {\bibinfo  {journal} {Astron. Astrophys.}\ }\textbf {\bibinfo {volume}
  {24}},\ \bibinfo {pages} {337} (\bibinfo {year} {1973})}\BibitemShut
  {NoStop}%
\bibitem [{\citenamefont {Potter}(2021)}]{Potter:2021vlg}%
  \BibitemOpen
  \bibfield  {author} {\bibinfo {author} {\bibfnamefont {W.~J.}\ \bibnamefont
  {Potter}},\ }\bibfield  {title} {\bibinfo {title} {{A full relativistic thin
  disc -- the physics of the plunging region and the value of the stress at the
  ISCO}},\ }\href {https://doi.org/10.1093/mnras/stab636} {\bibfield  {journal}
  {\bibinfo  {journal} {Mon. Not. Roy. Astron. Soc.}\ }\textbf {\bibinfo
  {volume} {503}},\ \bibinfo {pages} {5025} (\bibinfo {year} {2021})},\ \Eprint
  {https://arxiv.org/abs/2104.09951} {arXiv:2104.09951 [astro-ph.HE]}
  \BibitemShut {NoStop}%
\bibitem [{\citenamefont {Garc{\'\i}a}\ \emph {et~al.}(2016)\citenamefont
  {Garc{\'\i}a}, \citenamefont {Fabian}, \citenamefont {Kallman}, \citenamefont
  {Dauser}, \citenamefont {Parker}, \citenamefont {McClintock}, \citenamefont
  {Steiner},\ and\ \citenamefont {Wilms}}]{Garcia:2016wse}%
  \BibitemOpen
  \bibfield  {author} {\bibinfo {author} {\bibfnamefont {J.~A.}\ \bibnamefont
  {Garc{\'\i}a}}, \bibinfo {author} {\bibfnamefont {A.~C.}\ \bibnamefont
  {Fabian}}, \bibinfo {author} {\bibfnamefont {T.~R.}\ \bibnamefont {Kallman}},
  \bibinfo {author} {\bibfnamefont {T.}~\bibnamefont {Dauser}}, \bibinfo
  {author} {\bibfnamefont {M.~L.}\ \bibnamefont {Parker}}, \bibinfo {author}
  {\bibfnamefont {J.~E.}\ \bibnamefont {McClintock}}, \bibinfo {author}
  {\bibfnamefont {J.~F.}\ \bibnamefont {Steiner}},\ and\ \bibinfo {author}
  {\bibfnamefont {J.}~\bibnamefont {Wilms}},\ }\bibfield  {title} {\bibinfo
  {title} {{High-Density Effects in X-ray Reflection Models from Accretion
  Disks}},\ }\href {https://doi.org/10.1093/mnras/stw1696} {\bibfield
  {journal} {\bibinfo  {journal} {Mon. Not. Roy. Astron. Soc.}\ }\textbf
  {\bibinfo {volume} {462}},\ \bibinfo {pages} {751} (\bibinfo {year}
  {2016})},\ \Eprint {https://arxiv.org/abs/1603.05259} {arXiv:1603.05259
  [astro-ph.HE]} \BibitemShut {NoStop}%
\bibitem [{\citenamefont {Wu}\ \emph {et~al.}(2023)\citenamefont {Wu},
  \citenamefont {Chen},\ and\ \citenamefont {Lin}}]{Wu:2023qeh}%
  \BibitemOpen
  \bibfield  {author} {\bibinfo {author} {\bibfnamefont {Y.}~\bibnamefont
  {Wu}}, \bibinfo {author} {\bibfnamefont {Y.-X.}\ \bibnamefont {Chen}},\ and\
  \bibinfo {author} {\bibfnamefont {D.~N.~C.}\ \bibnamefont {Lin}},\ }\bibfield
   {title} {\bibinfo {title} {{Chaotic Type I migration in turbulent discs}},\
  }\href {https://doi.org/10.1093/mnrasl/slad183} {\bibfield  {journal}
  {\bibinfo  {journal} {Mon. Not. Roy. Astron. Soc.}\ }\textbf {\bibinfo
  {volume} {528}},\ \bibinfo {pages} {L127} (\bibinfo {year} {2023})},\ \Eprint
  {https://arxiv.org/abs/2311.15747} {arXiv:2311.15747 [astro-ph.EP]}
  \BibitemShut {NoStop}%
\bibitem [{\citenamefont {Chapman-Bird}\ \emph {et~al.}(2025)\citenamefont
  {Chapman-Bird} \emph {et~al.}}]{Chapman-Bird:2025xtd}%
  \BibitemOpen
  \bibfield  {author} {\bibinfo {author} {\bibfnamefont {C.~E.~A.}\
  \bibnamefont {Chapman-Bird}} \emph {et~al.},\ }\bibfield  {title} {\bibinfo
  {title} {{The Fast and the Frame-Dragging: Efficient waveforms for
  asymmetric-mass eccentric equatorial inspirals into rapidly-spinning black
  holes}},\ }\href@noop {} {\  (\bibinfo {year} {2025})},\ \Eprint
  {https://arxiv.org/abs/2506.09470} {arXiv:2506.09470 [gr-qc]} \BibitemShut
  {NoStop}%
\bibitem [{\citenamefont {Berti}\ and\ \citenamefont
  {Volonteri}(2008)}]{Berti:2008af}%
  \BibitemOpen
  \bibfield  {author} {\bibinfo {author} {\bibfnamefont {E.}~\bibnamefont
  {Berti}}\ and\ \bibinfo {author} {\bibfnamefont {M.}~\bibnamefont
  {Volonteri}},\ }\bibfield  {title} {\bibinfo {title} {{Cosmological black
  hole spin evolution by mergers and accretion}},\ }\href
  {https://doi.org/10.1086/590379} {\bibfield  {journal} {\bibinfo  {journal}
  {Astrophys. J.}\ }\textbf {\bibinfo {volume} {684}},\ \bibinfo {pages} {822}
  (\bibinfo {year} {2008})},\ \Eprint {https://arxiv.org/abs/0802.0025}
  {arXiv:0802.0025 [astro-ph]} \BibitemShut {NoStop}%
\bibitem [{\citenamefont {King}\ \emph {et~al.}(2008)\citenamefont {King},
  \citenamefont {Pringle},\ and\ \citenamefont {Hofmann}}]{King:2008au}%
  \BibitemOpen
  \bibfield  {author} {\bibinfo {author} {\bibfnamefont {A.~R.}\ \bibnamefont
  {King}}, \bibinfo {author} {\bibfnamefont {J.~E.}\ \bibnamefont {Pringle}},\
  and\ \bibinfo {author} {\bibfnamefont {J.~A.}\ \bibnamefont {Hofmann}},\
  }\bibfield  {title} {\bibinfo {title} {{The Evolution of Black Hole Mass and
  Spin in Active Galactic Nuclei}},\ }\href
  {https://doi.org/10.1111/j.1365-2966.2008.12943.x} {\bibfield  {journal}
  {\bibinfo  {journal} {Mon. Not. Roy. Astron. Soc.}\ }\textbf {\bibinfo
  {volume} {385}},\ \bibinfo {pages} {1621} (\bibinfo {year} {2008})},\ \Eprint
  {https://arxiv.org/abs/0801.1564} {arXiv:0801.1564 [astro-ph]} \BibitemShut
  {NoStop}%
\bibitem [{\citenamefont {Dyson}\ and\ \citenamefont
  {D'Orazio}()}]{DysonToAppear}%
  \BibitemOpen
  \bibfield  {author} {\bibinfo {author} {\bibfnamefont {C.}~\bibnamefont
  {Dyson}}\ and\ \bibinfo {author} {\bibfnamefont {D.~J.}\ \bibnamefont
  {D'Orazio}},\ }\bibfield  {title} {\bibinfo {title} {{\textit{To appear}}},\
  }\href@noop {} {\ }\BibitemShut {NoStop}%
\bibitem [{\citenamefont {Copparoni}\ \emph {et~al.}(2025)\citenamefont
  {Copparoni}, \citenamefont {Barausse}, \citenamefont {Speri}, \citenamefont
  {Sberna},\ and\ \citenamefont {Derdzinski}}]{Copparoni:2025jhq}%
  \BibitemOpen
  \bibfield  {author} {\bibinfo {author} {\bibfnamefont {L.}~\bibnamefont
  {Copparoni}}, \bibinfo {author} {\bibfnamefont {E.}~\bibnamefont {Barausse}},
  \bibinfo {author} {\bibfnamefont {L.}~\bibnamefont {Speri}}, \bibinfo
  {author} {\bibfnamefont {L.}~\bibnamefont {Sberna}},\ and\ \bibinfo {author}
  {\bibfnamefont {A.}~\bibnamefont {Derdzinski}},\ }\bibfield  {title}
  {\bibinfo {title} {{Implications of stochastic gas torques for asymmetric
  binaries in the LISA band}},\ }\href
  {https://doi.org/10.1103/PhysRevD.111.104079} {\bibfield  {journal} {\bibinfo
   {journal} {Phys. Rev. D}\ }\textbf {\bibinfo {volume} {111}},\ \bibinfo
  {pages} {104079} (\bibinfo {year} {2025})},\ \Eprint
  {https://arxiv.org/abs/2502.10087} {arXiv:2502.10087 [gr-qc]} \BibitemShut
  {NoStop}%
\end{thebibliography}%
\end{document}